\documentclass[twocolumn,showpacs,showkeys,preprintnumbers,amsmath,amssymb,prd,letterpaper,floatfix,superscriptaddress,10pt,aps,nofootinbib]{revtex4-1}
\usepackage[margin=1in,footskip=0.25in]{geometry}
\usepackage[utf8]{inputenc}
\usepackage{graphicx}
\usepackage{epsfig}
\usepackage{color}
\usepackage{longtable}
\usepackage[breaklinks=true]{hyperref}
\usepackage{latexsym}
\usepackage{amsmath}
\usepackage{amssymb}
\usepackage{dsfont}
\usepackage{verbatim}
\usepackage{bm}
\usepackage{bbm}
\usepackage{placeins}

\hypersetup{colorlinks, linkcolor = [rgb]{0,0.0,0.75}, citecolor = [rgb]{0,0.0,0.75}, urlcolor = [rgb]{0,0.0,0.75}}

\newcommand{\I}{\mathrm{i}}
\newcommand{\E}{\mathrm{e}}

\newcommand{\brackets}[1]{\langle #1 \rangle}
\newcommand{\bracketsf}[1]{\langle #1 \rangle_f}
\newcommand{\epow}[1]{\mathrm{e}^{ #1 }}
\newcommand{\Tr}[1]{\mathrm{Tr}\big( #1 \big)}
\newcommand{\tdagger}{ {\tilde{\dagger}}}

\newcommand{\ubar}{\bar{u}}
\newcommand{\dbar}{\bar{d}}
\newcommand{\qbar}{\bar{q}}
\newcommand{\psibar}{\bar{\psi}}
\newcommand{\pvec}{\vec{p}}
\newcommand{\xvec}{\vec{x}}
\newcommand{\refeq}[1]{(\ref{#1})}
\newcommand{\qqvec}{k_n^{\Lambda, \vec{P}},L}

\newcommand{\ape}{\mathrm{APE}}
\newcommand{\Expval}{\mathrm{E}}

\def\MIT{Center for Theoretical Physics, Laboratory for Nuclear Science and Department of Physics, Massachusetts Institute of Technology, Cambridge, Massachusetts 02139, USA}
\def\RFWUB{Helmholtz-Institut f\"ur Strahlen- und Kernphysik, Rheinische Friedrich-Wilhelms-Universit\"at Bonn, Nu{\ss}allee 14-16, D-53115 Bonn, Germany}
\def\CYI{Computation-based Science and Technology Research Center, Cyprus Institute, 20 Kavafi Str., 2121 Nicosia, Cyprus}
\def\UCY{Department of Physics, University of Cyprus, P.O. Box 20537, 1678 Nicosia, Cyprus}
\def\RIKEN{RIKEN BNL Research Center, Brookhaven National Laboratory, Upton, NY 11973, USA}

\def\UOA{Department of Physics, University of Arizona, Tucson, AZ 85721, USA}
\def\SBU{Department of Physics and Astronomy, Stony Brook University, Stony Brook, NY 11794, USA}

\begin{document}

\author{Constantia Alexandrou}
  \affiliation{\UCY}
  \affiliation{\CYI}

\author{Luka Leskovec}
\email{leskovec@email.arizona.edu}
  \affiliation{\UOA}

\author{Stefan Meinel}
\email{smeinel@email.arizona.edu}
  \affiliation{\UOA}
  \affiliation{\RIKEN}

\author{John Negele}
  \affiliation{\MIT}

\author{\mbox{Srijit Paul}}
  \affiliation{\CYI}

\author{Marcus Petschlies}
\email{marcus.petschlies@hiskp.uni-bonn.de}
  \affiliation{\RFWUB}

\author{Andrew Pochinsky}
  \affiliation{\MIT}

\author{Gumaro Rendon}
  \affiliation{\UOA}

\author{Sergey Syritsyn}
  \affiliation{\RIKEN}
  \affiliation{\SBU}


\date{July 20, 2017}
\title{\texorpdfstring{$\bm{P}$-wave $\bm{\pi\pi}$ scattering and the $\bm{\rho}$ resonance from lattice QCD}{}}

\vspace{0.2cm}

\begin{abstract}
We calculate the parameters describing elastic $I=1$, $P$-wave $\pi\pi$
scattering using lattice QCD with $2+1$ flavors of clover fermions.
Our calculation is performed with a pion mass of $m_\pi \approx 320\:\:{\rm MeV}$
and a lattice size of $L\approx 3.6$ fm.
We construct the two-point correlation matrices with both quark-antiquark and two-hadron
interpolating fields using a combination of smeared forward, sequential and stochastic propagators.
The spectra in all relevant irreducible representations for total momenta $|\vec{P}|
\leq \sqrt{3} \frac{2\pi}{L}$ are extracted with two alternative methods: a variational analysis
as well as multi-exponential matrix fits. We perform an analysis using L\"uscher's formalism
for the energies below the inelastic thresholds, and investigate several phase shift models,
including possible nonresonant contributions.
We find that our data are well described by the minimal
Breit-Wigner form, with no statistically significant nonresonant component.
In determining the $\rho$ resonance mass and coupling we compare two different
approaches: fitting the individually extracted phase shifts versus fitting the
$t$-matrix model directly to the energy spectrum. We find that both methods give
consistent results, and at a pion mass of $am_{\pi}=0.18295(36)_{stat}$ obtain $g_{\rho\pi\pi} = 5.69(13)_{stat}(16)_{sys}$,
$am_\rho = 0.4609(16)_{stat}(14)_{sys}$, and $am_{\rho}/am_{N} = 0.7476(38)_{stat}(23)_{sys} $, where
the first uncertainty is statistical and the second is the systematic uncertainty
due to the choice of fit ranges.
\end{abstract}

\maketitle

\section{Introduction}\label{sec_introduction}

One of the most fascinating phenomena of QCD is the hadronic spectrum: a complex set of
composite particles arising from the interactions between quarks and gluons. If we neglect
the electromagnetic and weak interactions, we can distinguish hadrons that are stable,
i.e.~those that do not decay via the strong interaction (for example the pion), and
hadrons that are unstable, such as the $\rho$ meson.

The $\rho$ meson is an isotriplet of short-lived hadronic resonances with quantum numbers
$J^{PC}=1^{--}$, which has been observed in multiple decay modes, including $\pi \pi$
(with a branching ratio of $99.9\%$), $\pi\pi\pi\pi$, $K\bar{K}$, and $\pi \gamma$
\cite{Olive:2016xmw}. The two most important parameters of the $\rho$ meson are its resonant mass
$m_{\rho}$ and its decay width $\Gamma_{\rho\to\pi\pi}$. Both have been studied
extensively with lattice QCD \cite{Gottlieb:1985rc,McNeile:2002fh,Aoki:2007rd,Gockeler:2008kc,
Jansen:2009hr, Feng:2010es, Frison:2010ws,Lang:2011mn,Aoki:2011yj,Pelissier:2012pi,
Dudek:2012xn,Wilson:2015dqa, Bali:2015gji,Bulava:2016mks,Hu:2016shf,Guo:2016zos,Fu:2016itp},
but many questions remain open, concerning for example the detailed dependence on the quark masses,
the effects of $N_f=2+1$ versus $N_f=2$ sea quarks, the coupling to the $K\bar{K}$ channel, and
the size of discretization errors for different lattice actions.

The $\rho$ resonance corresponds to a pole in the $I=1$ $P$-wave $\pi\pi$ scattering amplitude.
This scattering amplitude plays an important role in many Standard Model processes,
and its energy dependence must be determined accurately as part of lattice calculations of matrix elements involving
the $\rho$ \cite{Briceno:2014uqa}, such as $\pi \gamma \to \rho (\to \pi \pi)$ \cite{Briceno:2015dca, Briceno:2016kkp}
and $B \to \rho (\to \pi \pi) \ell\bar{\nu}_\ell$.

In this work, we use the L\"uscher method to study the $\rho$ resonance in $\pi\pi$
scattering with lattice QCD. The energy levels of a two-hadron system in a finite volume
are shifted by the interactions between the hadrons. These energy shifts are related to
the infinite-volume scattering matrix via the L\"uscher quantization condition
\cite{Luscher:1990ux}. The L\"uscher method was initially derived for the scattering of
spin-$0$ particles in the rest frame \cite{Luscher:1990ux}, and was extended to moving
frames for the case of scattering of two particles with equal mass in
Refs.~\cite{Rummukainen:1995vs,Kim:2005gf,Christ:2005gi}. Further generalizations to
coupled channels, particles of unequal mass, arbitary spin, and three-particle systems
were given in Refs.~\cite{Hansen:2012tf,Leskovec:2012gb,Gockeler:2012yj,
Briceno:2014oea,Briceno:2017tce}.
Other methods that have been used to study resonances are the Hamiltionian
effective field theory approach \cite{Hall:2013qba}, which is similar to the L\"uscher
method, the HALQCD approach \cite{HALQCD:2012aa}, where the Nambu-Bethe-Salpeter wave
function is calculated and used to determine a potential between two hadrons, and the
method of Refs.~\cite{McNeile:2002az, Alexandrou:2013ata,Alexandrou:2015hxa}, which uses a
perturbative interpretation of the mixing of nearby states.

We construct two-point correlation matrices with two different types of interpolating
fields: quark-antiquark interpolators, and two-pion-scattering interpolators. From these
correlation matrices, we extract the energy spectrum below the $K\bar{K}$ and
$\pi\pi\pi\pi$ thresholds using two different analysis methods: 1) the variational
approach, also known as the generalized eigenvalue problem, and 2), multi-exponential fits
directly to the correlation matrix. We carefully compare the results from both methods and
estimate the systematic uncertainties associated with the choice of the fit range.

In our L\"uscher analysis of the elastic $\pi\pi$ scattering, we again compare two
different methods: 1) mapping each individual energy level to a corresponding scattering
phase shift, and then fitting Breit-Wigner-like models to the results, and 2) fitting the
models for the $t$-matrix directly to the energy spectrum, as was proposed in
Ref.~\cite{Guo:2012hv}. In constructing the models, we also allow for a possible
nonresonant contribution.

Our calculation includes $N_f=2+1$ dynamical quark flavors, implemented with a
clover-improved Wilson action. We use a single
ensemble of gauge configurations on a $32^3
\times 96$ lattice with $a\approx 0.114$ fm, corresponding to
a large physical volume of $(3.6\:{\rm fm})^3\times (10.9\:{\rm fm})$.
The calculation is performed in the isospin limit with a light-quark mass corresponding to
a pion mass of approximately 320 MeV.

The paper is organized as follows: We begin by briefly reviewing
the continuum description of elastic $\pi\pi$ scattering in
Sec.~\ref{sec_about_rho}. Section \ref{sec_lattice} contains our
lattice parameters and includes an analysis of the pion dispersion relation.
Our choice of interpolating fields and the construction of the two-point
correlation matrices are described in Sec.~\ref{sec_ops}, and the analysis of
the energy spectrum is reported in Sec.~\ref{sec_spectrum}. The formalism
of the L\"uscher analysis is reviewed in Sec.~\ref{sec_formalism}, while the
numerical results for the scattering phase shifts and resonance parameters
are discussed in Sec.~\ref{sec_results}.In Sec.~\ref{sec_results} we also
present a detailed comparison with previous lattice calculations and discuss
systematic uncertainties. We conclude in Sec.~\ref{sec_conclusions}.

\section{\texorpdfstring{About $\pi\pi$ scattering}{}}\label{sec_about_rho}

In this section we briefly review the formalism describing elastic $\pi\pi$ $P$-wave
scattering in the $I(J^{PC}) = 1(1^{--})$ channel in the continuum \cite{Chung:1995dx}.

We express the $1\times1$ elastic scattering '`matrix'' as
\begin{align}
\label{eq:TfromS}
S_{\ell}(s) = 1 + 2 \I \;t_{\ell}(s),
\end{align}
where $t_\ell$ is the $t$-matrix (also known as the scattering amplitude), which depends
on the invariant mass $s$ of the system, and $\ell$ is the partial wave of the scattering channel.
The $t_\ell$ matrix is related to the scattering phase shift $\delta_\ell$ via
\begin{align}
t_{\ell}(s) = \frac{1}{\cot{\delta_{\ell}(s)}-\I}. \label{eq:tlofs}
\end{align}
A resonant contribution to $t_{\ell}(s)$ can be described\footnote{Note that a typical
Breit-Wigner model does not work for very broad resonance such as the $\sigma$ and
$\kappa$ scalar resonances \cite{Pelaez:2004vs}.} by a Breit-Wigner (BW) form,
\begin{align}
t_{\ell}(s) = \frac{\sqrt{s}\,\Gamma(s)}{m_R^2 - s - \I \sqrt{s}\,\Gamma(s)},
\end{align}
which corresponds to the phase shift
\begin{align}
\delta_\ell(s) = \arctan \frac{\sqrt{s}\,\Gamma(s)}{m_R^2 - s}.
\end{align}
In this work, we consider two different forms for the $\ell=1$ decay width $\Gamma(s)$:
\begin{itemize}

  \item {\bf BW I:} $P$-wave decay width:
  \begin{align}
  \label{eq:Gamma_Pwave}
  \Gamma_{I}(s) = \frac{g_{\rho\pi\pi}^2}{6\pi} \frac{k^{3}}{s},
  \end{align}
  where $g_{\rho\pi\pi}$ is the coupling between the $\pi\pi$ scattering channel and the
  $\rho$ resonance, and $k$ is the scattering momentum defined via $\sqrt{s} = 2
  \sqrt{m_{\pi}^2 + k^{2}}$. This form was used in most previous lattice QCD studies.

  \item {\bf BW II:} $P$-wave decay width modified with Blatt-Weisskopf barrier factors \cite{VonHippel:1972fg}:
  \begin{align}
  \label{eq:Gamma_PwaveBW}
  \Gamma_{II}(s) = \frac{g_{\rho\pi\pi}^2}{6\pi} \frac{k^{3}}{s}\: \frac{1 + (k_R r_0)^2}{1 + (k r_0)^2},
  \end{align}
  where $k_R$ is the scattering momentum at the resonance position and $r_0$ is the radius
  of the centrifugal barrier.
\end{itemize}

In certain cases, for example in $P$-wave $N\pi$ scattering, the phase shift is known to
receive both resonant and nonresonant (NR) contributions \cite{Long:2009wq}. We also allow
for this possibilty in our analysis of $\pi\pi$ scattering and write the full $P$-wave
phase shift as
\begin{align}
\delta_1(s) = \delta_1^{BW}(s) + \delta_1^{NR}(s).
\end{align}
We investigate three different models for a nonresonant background contribution
$\delta_1^{NR}$:
\begin{itemize}
  \item {\bf NR I:} a constant nonresonant phase $A$:
  \begin{align}
  \delta_1^{NR}(s) = A.
  \end{align}

  \item {\bf NR II:} a nonresonant phase depending linearly on $s$:
  \begin{align}
  \delta_1^{NR}(s) = A + B s,
  \end{align}
  where $A$ and $B$ are free parameters.

  \item {\bf NR III:} zeroth order nonresonant effective-range expansion (ERE):
  \begin{align}
  \delta_1^{NR}(s) = \mathrm{arccot} \frac{2 a_1^{-1}}{\sqrt{s - s^{thres}}},
  \end{align}
  where $a_1^{-1}$ is the inverse scattering length and $s^{thres} = 4 m_{\pi}^2$ is the
  $\pi\pi$ threshold invariant mass.

\end{itemize}

\section{Lattice parameters}\label{sec_lattice}

\subsection{Gauge Ensemble}\label{sec_ensembles}

The parameters of the lattice gauge-field ensemble are given in Table \ref{tab:lattice}.
The gluon action is a tadpole-improved tree-level Symanzik action \cite{Symanzik:1983pq,
Symanzik:1983dc, Symanzik:1983gh, Luscher:1985zq}. We use the same clover-improved Wilson
action \cite{Wilson:1974sk, Sheikholeslami:1985ij} for the sea and valence quarks. The
gauge links in the fermion action are smeared using one level of stout smearing
\cite{Morningstar:2003gk} with staple weight $\rho=0.125$ (the smearing smoothes out
short-distance fluctuations and alleviates instabilities associated with low quark
masses). The lattice scale reported in Table \ref{tab:lattice} was determined from the
$\Upsilon(2S)-\Upsilon(1S)$ splitting \cite{Davies:2009tsa,Meinel:2010pv} calculated with
NRQCD \cite{Lepage:1992tx} at the physical $b$-quark mass. The strange-quark mass is
consistent with its physical value as indicated by the '`$\eta_s$'' mass
\cite{Davies:2009tsa, Dowdall:2011wh}.


\begin{table}[htb!]
\begin{tabular}{|r|c|}
\hline
    &   \texttt{C13}    \cr
\hline
  $N_L^3 \times N_T$  &   $32^3 \times 96 $     \cr
  $\beta$             &   $6.1$    \cr
  $N_f$               &   $2+1$                 \cr
  $c_{sw}$            &   $1.2493097$    \cr
  $a m_{u,d}$         &   $-0.285$    \cr
  $a m_{s}$           &   $-0.245$    \cr
  $N_{config}$        &   $1041$     \cr
  $a$ [fm]            &   $0.11403(77)$         \cr
  $L$ [fm]            &   $3.649(25)$                 \cr
  $am_{\pi}$          &   $0.18295(36)$          \cr
  $am_{N}$          &   $0.6165(23)$          \cr
  $am_{\eta_s}$       &   $0.3882(19)$          \cr
  $m_{\pi}L$          &   $5.865(32)$           \cr
\hline
\end{tabular}
\caption{Details of the gauge-field ensemble. $N_L$ and $N_T$ denote the number of lattice
points in the spatial and time directions. The lattice spacing, $a$, was determined using
the $\Upsilon(2S)-\Upsilon(1S)$ splitting. The ensemble was generated with $N_f=2+1$ flavors of sea
quarks with bare masses $a m_{u,d}$ and $a m_{s}$, which lead to the given values of
$am_{\pi}$, $am_{N}$, and $am_{\eta_s}$. The $\eta_s$ is an artificial pseudoscalar $s\bar{s}$ meson
that can be used to tune the strange-quark mass \cite{Davies:2009tsa, Dowdall:2011wh}. The
uncertainties given here are statistical only.}
\label{tab:lattice}
\end{table}

\subsection{The pion mass and dispersion relation}\label{sec_pion}

\begin{figure}[htb!]
  \centering
  \includegraphics[width=0.95\columnwidth]{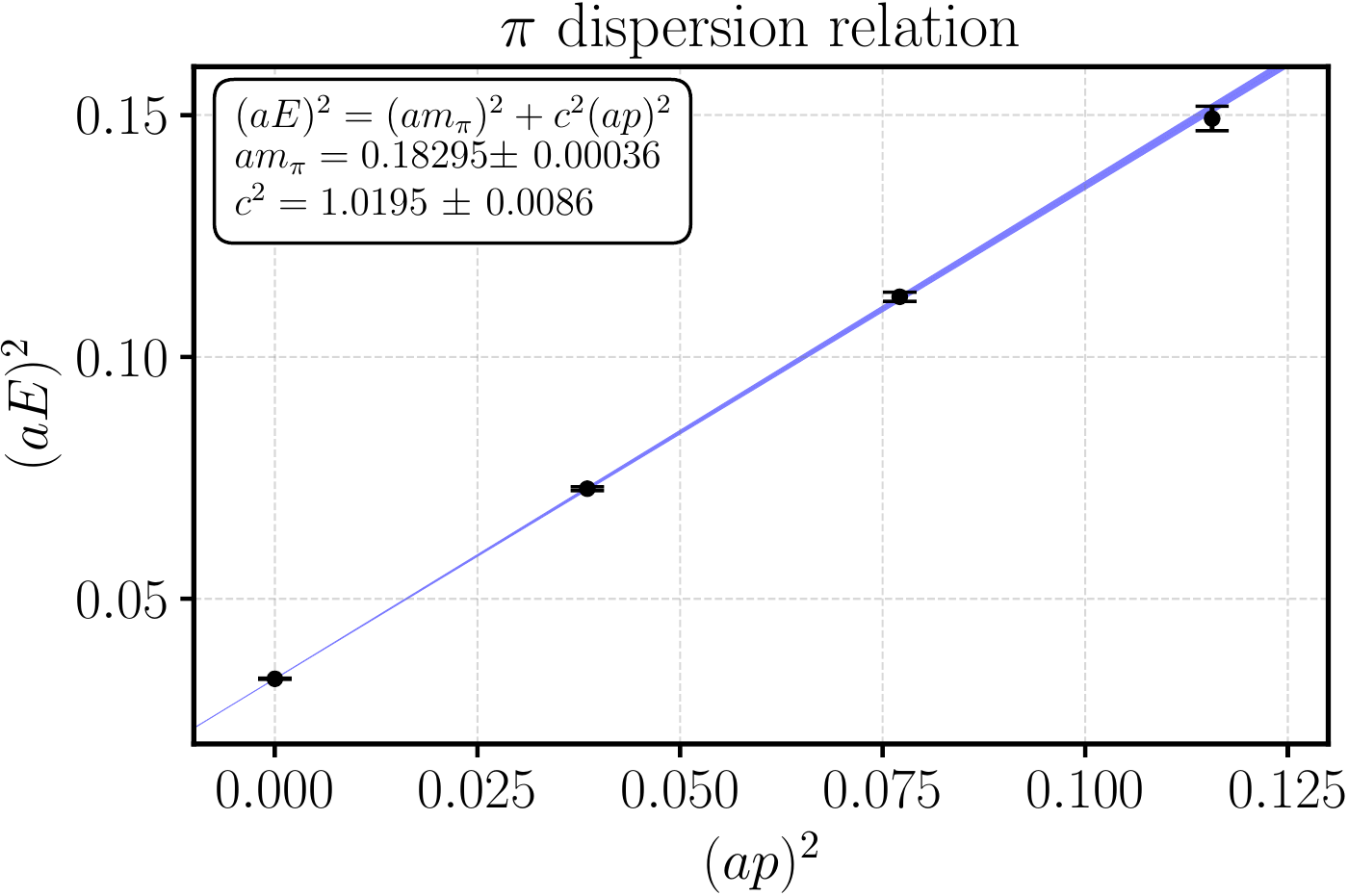}
  \caption{Pion dispersion relation. The $\pi$ mass and speed of light determined from the
  dispersion relation are consistent with a relativistic dispersion relation with the rest
  frame $\pi$ energy.}
  \label{fig:dispersion}
\end{figure}

To determine the $\rho$ resonance parameters with the L\"uscher method we need to know the
pion dispersion relation.  We performed a fit of the pion energies using the form $(aE)^2
= (am_{\pi})^2+c^2 (a p)^2$ in the range $0\leq p^2 \leq 3(2\pi/L)^2$, which yields
$am_\pi = 0.18295(36)$ and $c^2=1.0195(86)$, as shown in Fig.~\ref{fig:dispersion}. Given that $c^2$ is consistent with 1 within
2\%, we use the relativistic dispersion relation $(aE)^2 = (am_{\pi})^2+ (a p)^2$ in the
subsequent analysis.

\section{Interpolating fields and two-point functions}\label{sec_ops}

The L\"uscher quantization condition relates the infinite-volume $\pi\pi$ scattering
phase shifts to the finite-volume energy spectrum \cite{Luscher:1990ux}. The first step
in our calculation is therefore to determine this energy spectrum from appropriate
two-point correlation functions.

If there were no interactions between the two pions, the discrete energy levels of the
two-pion system in a cubic lattice of size $L$ would be equal to
\begin{align}
&E_{non-int}^{\vec{P}} = \sqrt{m_\pi^2 + |\vec{k}_1|^2} + \sqrt{m_\pi^2 + |\vec{k}_2|^2},
\end{align}
where
\begin{align}
& \vec{k}_1 = \frac{2\pi}{L} \vec{d}_1, \;\;   \vec{k}_2 = \frac{2\pi}{L} \vec{d}_2, \;\;  \vec{d}_1,\vec{d}_2 \in {\mathbb Z}^3,
\end{align}
and the total momentum is $\vec{P}=\vec{k}_1+ \vec{k}_2$. In the presence of interactions,
the individual momenta $\vec{k}_1$ and $\vec{k}_2$ are no longer good quantum numbers, but
the total momentum still is, and takes on the values
\begin{equation}
 \vec{P}= \frac{2\pi}{L} \vec{d}, \;\;  \vec{d} \in {\mathbb Z}^3.
\end{equation}
We denote the interacting energy levels as
\begin{equation}
E_n^{\vec{P}},
\end{equation}
where $n$ denotes the $n$-th state with the given total momentum (and any other relevant
quantum numbers). We relate these energies to the corresponding center-of-mass energies
\begin{align}
\label{eq:invmass}
E_{n,\,CM}^{\vec{P}} = \sqrt{s_n^{ \vec{P}}}  = \sqrt{(E_n^{ \vec{P}})^2 - \vec{P}^2},
\end{align}
and define the scattering momentum $k_n^{ \vec{P}}$ via
\begin{align}
\label{eq:scattmom}
\sqrt{s_n^{ \vec{P}}} = 2 \sqrt{m_{\pi}^2 + (k_n^{ \vec{P}})^2}.
\end{align}
Note that $k_n^{ \vec{P}}$ is not a lattice momentum, and can take on continuous (possibly
even imaginary) values. The interacting energy levels, and hence the scattering momenta,
depend on the scattering phase shifts, the lattice size $L$, and the symmetries of the
two-particle system, as described by the L\"uscher quantization condition and its
generalization to moving frames \cite{Luscher:1990ux,Kim:2005gf,Christ:2005gi}.

We aim to determine the values of the scattering phase shift $\delta_1(s)$ for many values
of $s$ near the $\rho$ resonance mass. The fairly large lattice volume we use
($L\approx 3.6$ fm) allows us to obtain a sufficient number of energy levels in the region of
interest from only the single volume combined with multiple moving frames, $\vec{P}$. In
this work, we use the moving frames and irreducible representations ($\Lambda$) listed in
Table \ref{tab:irreps}.

\begin{table}[htb!]
\begin{tabular}{|c|c|c|c|}
\hline
  $\vec{P}$ $[\frac{2\pi}{L}]$ &   Little Group & Irrep $\Lambda$  & $J$ \cr
\hline
  $(0,0,0)$  &  $O_h$   &  $T_1^-$ &  $1^-,3^-,\ldots$\cr
  $(0,0,1)$  & $D_{4h}$ (${\rm Dic}_4$) &  $A_2^-$ ($A_1$) &  $1^-,3^-,\ldots$\cr
  $(0,0,1)$  & $D_{4h}$ (${\rm Dic}_4$) &  $E^-$   ($E$)   &  $1^-,3^-,\ldots$\cr
  $(0,1,1)$  & $D_{2h}$ (${\rm Dic}_2$) &  $B_1^-$ ($A_1$) &  $1^-,3^-,\ldots$\cr
  $(0,1,1)$  & $D_{2h}$ (${\rm Dic}_2$) &  $B_2^-$ ($B_1$) &  $1^-,3^-,\ldots$\cr
  $(0,1,1)$  & $D_{2h}$ (${\rm Dic}_2$) &  $B_3^-$ ($B_2$) &  $1^-,3^-,\ldots$\cr
  $(1,1,1)$  & $D_{3d}$ (${\rm Dic}_3$) &  $A_2^-$ ($A_1$) &  $1^-,3^-,\ldots$\cr
  $(1,1,1)$  & $D_{3d}$ (${\rm Dic}_3$) &  $E^-$   ($E$)   &  $1^-,3^-,\ldots$\cr
\hline
\end{tabular}
\caption{ The reference frames (i.e., total momenta $\vec{P}$), associated Little Groups,
and irreducible representations used to determine the multi-hadron spectrum in the
$I(J^{PC}) = 1(1^{--})$ channel. For the Little Groups and irreps with give both the Sch\"onflies
notation and the subduction notation. Due to a reduction in symmetry, the Little Group irreps
$\Lambda$ contain not only $J^P=1^{-}$ states, but also higher $J$, starting with $J=3$.
In the channel we investigate, the $J=3$ contributions have been shown to be negligible
\cite{Estabrooks:1975cy, Dudek:2012xn}.}
\label{tab:irreps}
\end{table}

\subsection{Interpolating fields}\label{subsec_operators}

The spectra in the frames and irreps listed in Table \ref{tab:irreps} are obtained from
two-point correlation functions constructed using two different types of interpolating
fields: local single-hadron quark-antiquark interpolating fields $\big\{ O_{\qbar q}
\big\}$, and two-hadron interpolating fields $\big\{ O_{\pi\pi} \big\}$. We choose the
quantum numbers $J^{PC}=1^{--}$ and $I=1,I_3=1$ (corresponding to the $\rho^+$
resonance\footnote{Due to the exact isospin symmetry in our lattice QCD calculation all
three isospin components $\rho^+, \rho^-$, and $\rho^0$ have the same properties.}), and
write
\begin{align}
  \label{eq:Oqq}
  O_{\qbar q} \big(t , \vec{P}\big) &=
  \sum_{\xvec} \, \bar{d}(t,\xvec)\,\Gamma \,u(t, \xvec) \,\epow{\I\vec{P}\cdot\xvec}\,,\\
    \nonumber  \label{eq:Opipi}
  O_{\pi\pi}\big(t, \pvec_1, \pvec_2\big) &=
    \frac{1}{\sqrt{2}} \,\big(\pi^+(t,\pvec_1)\, \pi^0(t,\pvec_2)\\
     &\qquad- \pi^0(t,\pvec_1)\,\pi^+(t,\pvec_2) \big)\, ,
\end{align}
where $\vec{P}=\vec{p}_1+\vec{p}_2$, and the single-pion interpolators are given by
\begin{align*}
  \pi^+(t,\pvec)  &= \sum_{\xvec}\dbar(t,\xvec)\,\gamma_5\,u(t,\xvec)\,\epow{i\pvec\cdot\xvec} \\
  \pi^0(t,\pvec)  &= \sum_{\xvec}\frac{1}{\sqrt{2}} \,\big(
    \ubar(t,\xvec)\,\gamma_5\,u(t,\xvec)\\
    \nonumber
    &\qquad -\dbar(t,\xvec)\,\gamma_5\,d(t,\xvec)
  \big) \,\epow{i\pvec\cdot\xvec}\, .
\end{align*}
We do not include quark-antiquark interpolators with derivatives, as past calculations
have shown that such interpolators do not improve the determination of the spectrum near
the $\rho$ resonance mass region \cite{Lang:2011mn}.

In Eq.~(\ref{eq:Oqq}), we use two different $\Gamma_i$ matrices, namely $\gamma_i$ and
$\gamma_0\gamma_i$, to obtain overlap with the $I(J^{PC})=1(1^{--})$ quantum numbers. The
single-hadron interpolators are projected to the finite-volume irreps $\Lambda$ of the
Little Group $LG(\vec{P})$ for the momentum $\vec{P}$ using
\begin{equation}
 O_{\qbar q}^{\Lambda,\,\vec{P}}(t) = \frac{\mathrm{dim}(\Lambda)}{N_{LG(\vec{P})}}
\sum_{ \hat{R} \in LG(\vec{P}) } \chi_\Lambda(\hat{R}) \hat{R}\,O_{\qbar q}(t,\vec{P}),
\end{equation}
where $\mathrm{dim}(\Lambda)$ is the dimension of the irrep, $N_{LG(\vec{P})}$ is the
order of the Little Group, and $\chi_\Lambda(\hat{R})$ is the character of $\hat{R} \in
LG(\vec{P})$ \cite{Dresselhaus:2008}.

The second interpolator type, Eq.~(\ref{eq:Opipi}), is built from products of two
single-pion interpolators, each separately projected to a definite momentum.
In this case, the projection proceeds through the formula given in
Ref.~\cite{Feng:2010es}:
\begin{align}
\label{eq:projection_operator}
&O_{\pi\pi}^{\Lambda,\,\vec{P}}(t) = \frac{\mathrm{dim}(\Lambda)}{N_{LG(\vec{P})}}
\sum_{ \hat{R} \in LG(\vec{P}) }\cr
&\chi_\Lambda(\hat{R}) \bigg( \pi^+(t,\vec{P}/2+\hat{R}\vec{p}\,)\:\pi^0(t,\vec{P}/2-\hat{R}\vec{p}\,)\cr
&\hspace{3ex} -\pi^0(t,\vec{P}/2+\hat{R}\vec{p}\,)\: \pi^+(t,\vec{P}/2-\hat{R}\vec{p}\,) \bigg),
\end{align}
where
\begin{align}
 \vec{p}=\frac{\vec{P}}{2} + \frac{2\pi}{L}\vec{m}, \;\;\vec{m}\in\mathbb{Z}^3.
\end{align}
(An alternative method to construct the interpolators is the subduction method
\cite{Moore:2005dw,Dudek:2012gj,Prelovsek:2016iyo}, which gives the same types of
interpolators as we find with the projection method.)

In the following, we use the schematic notation $O_1$ for quark-antiquark interpolators
with $\gamma_i$, $O_2$ for quark-antiquark interpolators with $\gamma_0\gamma_i$, and
$O_3$, $O_4$ for two-pion interpolators with the smallest and second-smallest possible
$\vec{p}$ in the given irrep.

\subsection{Wick contractions}\label{subsec_wick}

\begin{figure}[!htb]
  \centering
  \includegraphics[width=0.47\textwidth]{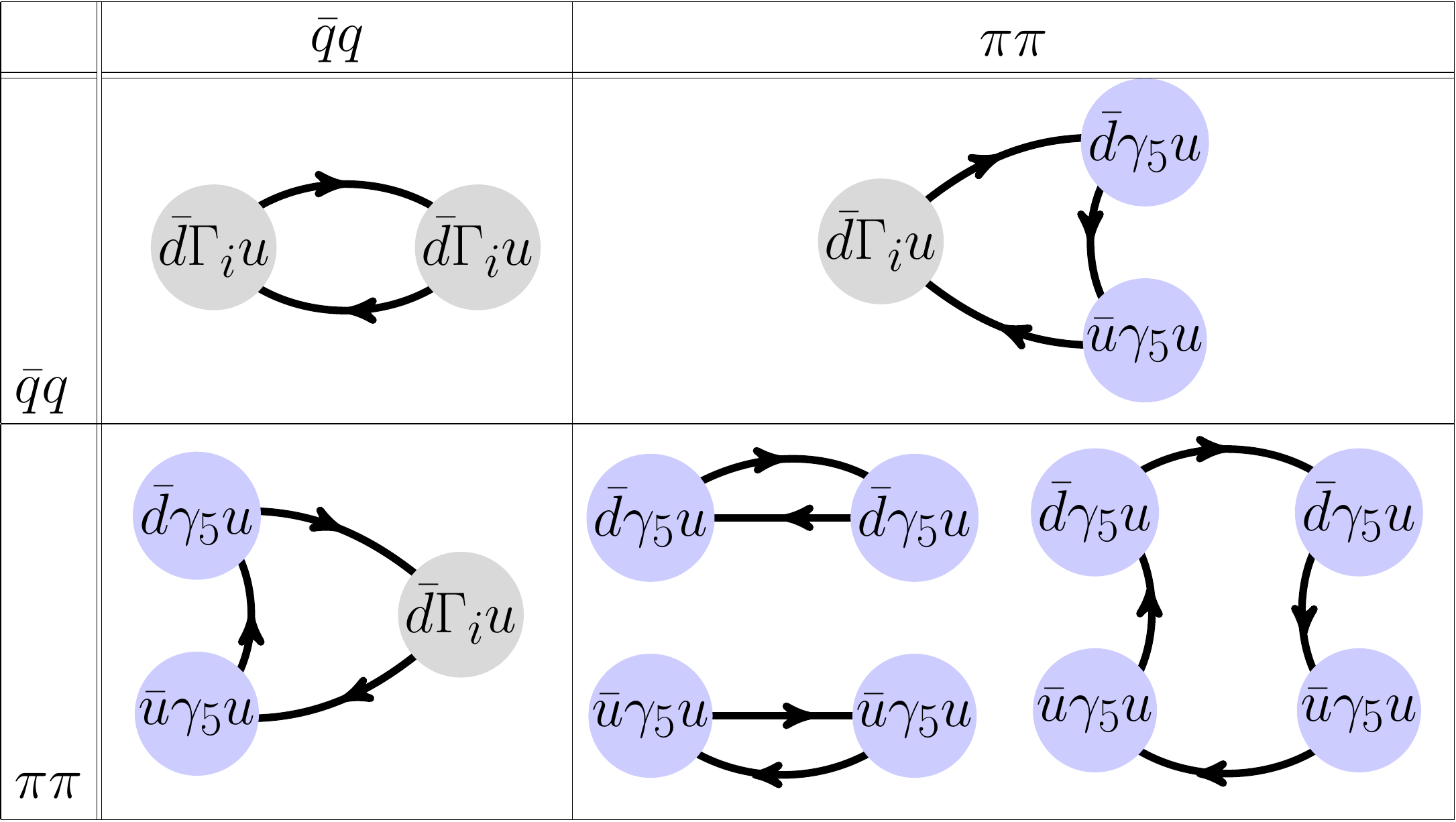}
  \caption{\label{fig:wick2pt} The Wick contractions corresponding to the correlation matrix elements of type $C_{\qbar q - \qbar q}$,
  $C_{\pi\pi - \qbar q}$, $C^\mathrm{direct}_{\pi\pi - \pi\pi}$ and $C^\mathrm{box}_{\pi\pi - \pi\pi}$.}
\end{figure}

The correlation matrix $C_{ij}^{\Lambda,\vec{P}}(t)$ is obtained from the interpolators
defined above as
\begin{align}
  C_{ij}^{\Lambda,\vec{P}}(t_f - t_i) &= \brackets{O_i^{\Lambda,\vec{P}}(t_f)\: O_j^{\Lambda,\vec{P}}(t_i)^\dagger}\,,
  \label{eq:correlation_matrix_element}
\end{align}
where $t_i$ is the source time and $t_f$ is the sink time. The correlation matrix elements
are expressed in terms of quark propagators by performing the Wick contractions (i.e., by
performing the path integral over the quark fields in a given gauge-field configuration).
The resulting quark-flow diagrams are shown in Fig.~\ref{fig:wick2pt} (for the case $I=1$
considered here, further disconnected diagrams cancel due to exact isospin symmetry). In
this section, we use the generic notation $\qbar q$ for the $i=1,2$ interpolators and
$\pi\pi$ for the $i=3,4$ interpolators to describe our method.

The diagrams in Fig.~\ref{fig:wick2pt} are obtained from point-to-all propagators (labeled
$f$), sequential propagators (labeled $seq$) and stochastic timeslice-to-all propagators
(labeled $st$). In detail, these propagator types are given as follows:

\paragraph*{a. Point-to-all propagator:}
Writing the quark and anti-quark fields as $\psi(t_f,
\xvec)_{\alpha}^{a}$ and $\bar{\psi}(t_i, \xvec_i)_{\beta}^{b}$, where $\alpha,\beta$ are
spin indices and $a,b$ are color indices, the point-to-all propagator $S_f$ from the fixed
initial point $x_i = (t_i, \xvec_i)$ to any final point $x_f = (t_f, \xvec_f)$ on the
lattice is the matrix element of the inverse of the lattice Dirac operator $D$:

\vspace{2ex}

\begin{minipage}[htb]{0.45\textwidth}
  \begin{center}
  \includegraphics[width=0.55\textwidth]{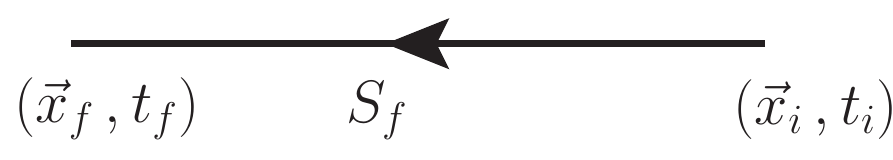}
  \end{center}
\end{minipage}

\begin{center}
\begin{align}
\nonumber  &S_f(t_f, \xvec; t_i, \xvec_i)_{\alpha\beta}^{ab} = \bracketsf{\psi(t_f,\xvec_f)^a_\alpha\,\psibar(t_i,\xvec_i)^b_\beta} \\
  &= D^{-1}(t_f,\xvec_f; t_i, \xvec_i)_{\alpha\beta}^{ab}\,.
\end{align}
\end{center}

\paragraph*{b. Sequential propagator:}
The sequential propagator describes the quark flow through a vertex of a given flavor and
Lorentz structure. It is obtained from a point-to-all propagator by a second (sequential)
inversion on a source built from the point-to-all propagator with an inserted vertex at
timeslice $t_{seq}$ with spin structure $\Gamma$ and momentum insertion $\pvec$:\\

\begin{minipage}[htb]{0.45\textwidth}
  \begin{center}
    \includegraphics[width=0.55\textwidth]{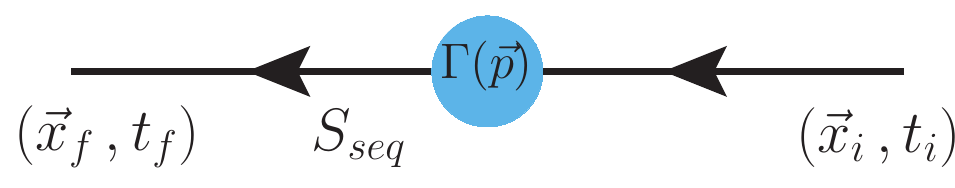}
  \end{center}
\end{minipage}

\begin{align}
  \label{eq:sequential_propagator}
 \nonumber &S_{seq}(t_f, \xvec_f; t_{seq},\pvec,\Gamma; t_i, \xvec_i) \\
  \nonumber
  &= \sum\limits_{\xvec_{seq}}\,D^{-1}(t_f,\xvec_f; t_{seq}, \xvec_{seq}) \\
  & \hspace{4ex} \times\,\Gamma\,\epow{i\vec{p}\cdot\vec{x}_{seq}}\,
  S_f(t_{seq}, \xvec_{seq}; t_i, \xvec_i)\,.
\end{align}

\paragraph*{c. Stochastic timeslice-to-all propagator:}

The stochastic timeslice-to-all propagator is defined as the inversion of the Dirac matrix
with a stochastic timeslice momentum source:

\vspace{2ex}

\begin{minipage}[htb]{0.45\textwidth}
  \begin{center}
    \includegraphics[width=0.55\textwidth]{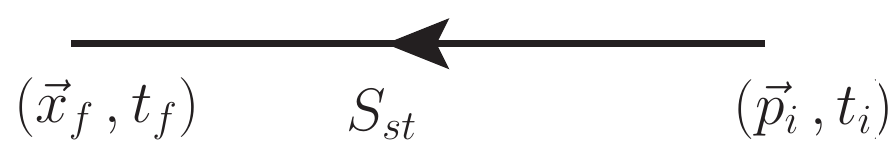}
  \end{center}
\end{minipage}
\begin{align}
 \nonumber \label{eq:S_st}
  &S_{st}(t_f,\xvec_f; t_i,\xvec_i) =\\
  &\frac{1}{N_\mathrm{sample}}\,\sum\limits_{r=1}^{N_\mathrm{sample}}\,
  \phi^r_{t_i,\pvec_i}(t_f,\xvec_f)\,\xi^r_{t_i,\vec{0}}(t_i,\xvec_i)^\dagger\,,
\end{align}
where
\begin{align*}
  \phi^r_{t_i, \pvec_i} &= D^{-1}\,\xi^r_{t_i,\pvec_i} \, \mathrm{~and~}
  \xi^r_{t_i,\pvec_i}(t,\xvec) = \delta_{t,t_i}\,\epow{i\pvec_i\cdot\xvec }\, \xi^r_{t_i}(\xvec)\,.
\end{align*}
For each $r = 1,\dotsc,N_\mathrm{sample}$, $\xi^r_{t_i}$ is a spin-color timeslice vector
with independently distributed entries for real and imaginary part,
$\xi^r_{t_i}(t,\xvec)^a_\alpha \sim \mathbb{Z}_2 \times i\mathbb{Z}_2$, so that
the expectation values with respect to the stochastic
noise, denoted as $\Expval\big[ \quad \big]$, satisfy
\begin{align}
  \label{eq:Z2noise}
  &\Expval\big[
  \xi^r_{t_i}(t,\xvec)^a_\alpha
\big] = 0, \\
\nonumber
  &\Expval\big[
  \xi^{r_1}_{t_{i_1}}(\xvec_1)^{a_1}_{\alpha_1} \,
  \big( \xi^{r_2}_{t_{i_2}}(\xvec_2)^{a_2}_{\alpha_2} \big)^*
\big] &\\
&= \delta^{r_1,r_2}\,
  \delta_{t_{i_1},t_{i_2}} \,
  \delta_{\xvec_1, \xvec_2}\,,\delta_{\alpha_1,\alpha_2}\,\delta^{a_1,a_2}\,. \label{eq:Z2noise2}
\end{align}
This technique provides a good way to efficiently evaluate the box (and box-like) diagrams
with reasonable cost. In addition to time-dilution of the stochastic momentum source, we
also apply spin-dilution to make use of the efficient one-end-trick \cite{McNeile:2006bz}
in our contractions. In this case the stochastic sources read
\begin{align}
  \xi^r_{t_i,\pvec_i,\alpha}(t,\xvec)^b_\beta &= \delta_{t,t_i}\,\delta_{\alpha,\beta}\,
  \epow{i\pvec_i\cdot\xvec}\,\xi^r_{t_i}(\xvec)^b,
    \label{eq:S_st_oet}
\end{align}
and the color timeslice vectors $\xi^r_{t_i}$ have expectation values analoguous to those
in Eqs.~\refeq{eq:Z2noise} and \refeq{eq:Z2noise2}.

\paragraph*{d. Smearing:}

To enhance the dominance of the lowest lying states contributing to a correlator we apply
source and sink smearing to the propagator types listed above: for all inversions of the
Dirac matrix we replace $D^{-1} \to W\big[ U_{\ape} \big]\, D^{-1}\,W\big[ U_\ape
\big]^\dagger$, where $W\big[ U_\ape \big]$ denotes the Wuppertal-smearing operator
\cite{Gusken:1989ad} using an APE-smeared gauge field \cite{Albanese:1987ds} with the
parameters $n=25$, $\alpha_{APE}=2.5$. Since the source and sink smearing is always
understood, we will not denote it explicitly.

\paragraph*{e. Coherent sequential sources:}
In order to increase the available statistics for a fixed number of gauge configurations
we calculate all correlators for 8 equidistant source locations separated in time by
$T/8$ and with spatial source coordinates independently and uniformly sampled over the
spatial lattice. We then take results from all source locations and average over them.

To reduce the computational cost for the sequential propagators, we
insert 2 point-to-all propagators into a single sequential source before inverting the
Dirac matrix on the latter:
\begin{align}
  \label{eq:coherent_sequential_source}
  &S_{seq} = D^{-1}\,\xi_{seq}, \\
  \nonumber
  &\xi_{seq}(t,\xvec)  =
  \Gamma\,\epow{i\pvec\cdot\xvec} \,\bigg(
  \delta_{t,t^{(0)}_i} \, S_f \big( t^{(0)}_i,\xvec; t^{(0)}_i,\xvec_i \big)\\
   &+ \delta_{t,t^{(1)}_i} \, S_f \big( t^{(1)}_i,\xvec; t^{(1)}_i,\xvec_i \big) \bigg),
\end{align}
where $t^{(1)}_i = t^{(0)}_i + T / 2 \mathrm{~mod~} T$.\\

The correlation matrix is then built from the propagators listed above as follows:

\paragraph*{a. $\qbar q - \qbar q$ correlators:}
The typical 2-point correlator with a single-hadron interpolator at source and sink is
constructed using point-to-all propagators:
\begin{align}
  \label{eq:C_qbarq_qbarq}
  \nonumber &C_{\qbar q - \qbar q}(t_f - t_i; \pvec_{f},\Gamma_f; \pvec_{i},\Gamma_{i}) =\\
  \nonumber
  &-\sum_{\vec{x}_{f}}\,\Tr{
  (\gamma_5\,S_f(t_f,\xvec_f; t_i,\xvec_i) \,\gamma_5)^\tdagger \\
  &  \quad\times\Gamma_{f}\,S_f(t_f,\xvec_f; t_i,\xvec_i)\,\Gamma_{i}
}
\epow{i\pvec_{f}\cdot\xvec_f + i\pvec_i\cdot\xvec_i}\,.
\end{align}
Above, $(\quad)^\tdagger$ denotes the Hermitian adjoint with
respect to only spin-color indices. We use the convention $\vec{p}_f = -\vec{p}_i$.

The direct diagram of the $C_{\pi\pi - \pi\pi}$ correlation function is the product of two
of the previous correlators with $\Gamma_i = \gamma_5 = \Gamma_f$. However, translational
invariance allows only one of the $\xvec_i$ to be fixed. To perform the sum over
$\xvec_i$, we use the one-end-trick and define
\begin{align}
\nonumber & C_{\qbar q - \qbar q, oet}(t_{f} - t_{i}; \Gamma_f,\pvec_{f}; \Gamma_i,\pvec_{i}) =
  \label{eq:C_qbarq_qbarq_oet}\\
  \nonumber
  &-\sum\limits_{\alpha,\beta}\,\sum_{\xvec_{f}}(\Gamma_{i}\gamma_5)_{\alpha \beta}\,\phi_{t_{i},0,\beta}(t_f, \xvec_{f})^\tdagger\,
  \gamma_5\,\Gamma_{f}\\
  &\qquad \times \phi_{t_{i},\pvec_{i},\alpha}(t_f, \xvec_{f})\,
  \epow{i\pvec_{f}\cdot\xvec_{f}}\,,
\end{align}
where $\phi_{t_{i},0,\beta}$ and $\phi_{t_{i},\pvec_i,\alpha}$ are the spin-diluted
stochastic timeslice-to-all propagators from Eqs.~\refeq{eq:S_st} and \refeq{eq:S_st_oet}.
The stochastic-sample index $r$ is suppressed for brevity.

\paragraph*{b. $\pi\pi - \qbar q$ correlators:}
The only contribution to the $I=1$ correlators with a two-pion interpolator at the source
and a single-hadron interpolator at the sink reads
\begin{align}
\label{eq:C_pipi_qbarq}
\nonumber & C_{\qbar q - \pi\pi}(t_{f} - t_{i}; \Gamma_f, \pvec_{f}; \pvec_{i_1},\pvec_{i_2}) = \\
\nonumber
&-\sum\limits_{\xvec_{f}}\,
\Tr{
  S_f(t_f, \xvec_{f}; t_{i},\xvec_{i_1})^\tdagger\,\gamma_5\, \Gamma_{f}\\
  &\times S_{seq}(t_f,\xvec_{f}; t_{i},\pvec_{i_2}; t_{i},\xvec_{i_1})
}\,
\epow{i\pvec_{f}\cdot\xvec_{f} + i\pvec_{i_1}\cdot\xvec_{i_1}}\,,
\end{align}
where $S_{seq}$ is the sequential propagator from Eq.(\ref{eq:sequential_propagator}).

\paragraph*{c. $\pi\pi - \pi\pi$ correlators:}

The direct diagram in the lower right panel of Fig.~\ref{fig:wick2pt} is obtained as
the product of two $\qbar q - \qbar q$ correlators as
\begin{align}
\nonumber &  C^\mathrm{direct}_{\pi\pi - \pi\pi}(t_f-t_i; \pvec_{f_1},\pvec_{f_2},\pvec_{i_1},\pvec_{i_2})
  \label{eq:C_pipi_pipi_direct} \\
  \nonumber  & =  C_{\qbar q - \qbar q}(t_{f} - t_{i}; \gamma_5,\pvec_{f_1}; \gamma_5, \pvec_{i_1})\\
  &\quad\; \times C_{\qbar q - \qbar q, oet}(t_{f} - t_{i}; \gamma_5,\pvec_{f_2}; \gamma_5, \pvec_{i_2})\,.
\end{align}
The box-type diagram in the lower right panel of Fig.~\ref{fig:wick2pt} requires point-to-all,
 sequential, and stochastic propagators and is calculated in two steps:
\begin{align}
\nonumber  & C^\mathrm{box}_{\pi\pi - \pi\pi}(t_f - t_i, \pvec_{f_1},\pvec_{f_2},\pvec_{i_1},\pvec_{i_2})=
  \label{eq:box_diagram} \\
  \nonumber
  &-\frac{1}{N_\mathrm{sample}}\, \sum\limits_{r=1}^{N_\mathrm{sample}}\,
  \sum\limits_{\alpha,a}\,
  \eta^r_\phi\big(t_{f},t_{i};\pvec_{f_1};\xvec_{i_1} \big)^a_{\alpha} \\
  &\qquad \times \eta^r_\xi\big(t_{f},t_{i};\pvec_{f_2},\pvec_{i_2};\xvec_{i_1} \big)^a_{\alpha} \,
  \epow{i\pvec_{i_1}\cdot\xvec_{i_1}} \,,
\end{align}
where
\begin{align}
  \nonumber
  \label{eq:eta_xi}
  &\eta^r_\xi\big(t_{f},t_{i};\pvec_{f_2},\pvec_{i_2};\xvec_{i_1}\big) =
  \sum\limits_{\xvec_{f_2}}\,
  \xi^r_{t_{f} }(t_f, \xvec_{f_2})^\tdagger\,\gamma_5\\
  & \times S_{seq}\big(t_f, \xvec_{f_2}; t_{i},\pvec_{i_2}; t_i, \xvec_{i_1} \big)  \,
  \epow{i\pvec_{f_2}\cdot\xvec_{f_2}}
\end{align}
and
\begin{align}
  \nonumber
  \label{eq:eta_phi}
  &\eta^r_\phi\big(t_{f},t_{i};\pvec_{f_1};\xvec_{i_1}\big) =
  \sum\limits_{\xvec_{f_1}}\,
  S_f(t_f, \xvec_{f_1} ; t_i, \xvec_{i_1})^\tdagger\\
  &\times \phi^r_{t_{f},0}(t_f, \xvec_{f_1})\,
  \epow{i\pvec_{f_1}\cdot\xvec_{f_1}}\,.
\end{align}
In Eqs.~\refeq{eq:box_diagram}, \refeq{eq:eta_xi} and \refeq{eq:eta_phi} we used
$\gamma_5$-Hermiticity of the quark propagator as well as $\Gamma_{i_{1/2}} = \gamma_5 =
\Gamma_{f_{1/2}}$.

The $\pi\pi$-$\pi\pi$ elements of the correlation matrix are constructed as
\begin{align}
\nonumber &C_{\pi\pi - \pi\pi}(t_f-t_i; \pvec_{f_1},\pvec_{f_2},\pvec_{i_1},\pvec_{i_2}) = \\
\nonumber &\qquad\frac{1}{2}\,C^\mathrm{direct}_{\pi\pi - \pi\pi}(t_f-t_i; \pvec_{f_1},\pvec_{f_2},\pvec_{i_1},\pvec_{i_2}) \\
&\qquad -C^\mathrm{box}_{\pi\pi - \pi\pi}(t_f-t_i; \pvec_{f_1},\pvec_{f_2},\pvec_{i_1},\pvec_{i_2}).
\end{align}

\section{Spectrum results}\label{sec_spectrum}

We extract the energy levels $E_n^{\Lambda,\,\vec{P}}$ from the correlation matrices using
two alternative methods. The first method, discussed in Sec.~\ref{subsec_gevp}, is the
variational analysis, also known as the generalized eigenvalue problem (GEVP). The second
method, discussed in Sec.~\ref{subsec_MFA}, employs multi-exponential fits directly to the
correlation matrix.

\subsection{Variational analysis}\label{subsec_gevp}

\begin{table*}
\begin{tabular}{|c l l | c c c | l l l| c |}
\hline
$\frac{L}{2\pi}|\vec{P}|$ & $\Lambda$ & Basis & \hspace{0.5ex}  $n$ \hspace{0.5ex} & Fit range & \hspace{1ex} $\frac{\chi^2}{{\rm dof}}$ \hspace{0.5ex} & $aE_n^{\Lambda,\, \vec{P}} $ & $a\sqrt{s_n^{\Lambda,\, \vec{P}}}$ & $\delta_1$ $[^{\circ}]$ & Included   \cr
\hline
  $  0$ & $T_1$ & $O_{1234}$ & $1$ & $ 8$-$18$ & $0.82$ & $0.4588(16)(12)$  & $0.4588(16)(12)$  & $86.0(1.6)(1.2)$    &  Yes  \cr
  $  0$ & $T_1$ & $O_{1234}$ & $2$ & $ 8$-$18$ & $0.66$ & $0.5467(16)(9)$   & $0.5467(16)(9)$   & $166.5(2.1)(1.3)$   &  Yes  \cr
  $  0$ & $T_1$ & $O_{1234}$ & $3$ & $ 7$-$15$ & $1.54$ & $0.6713(41)(104)$ & $0.6713(41)(104)$ & $172.9(4.7)(168.1)$ &  No  \cr
\hline
  $  1$ & $A_2$ & $O_{1234}$ & $1$ & $ 8$-$18$ & $0.61$ & $0.44536(73)(23)$ & $0.39974(82)(25)$ & $2.81(25)(9)$ &  Yes  \cr
  $  1$ & $A_2$ & $O_{1234}$ & $2$ & $ 8$-$18$ & $1.04$ & $0.5124(20)(17)$ & $0.4732(22)(18)$ & $131.3(1.9)(1.6)$ &  Yes  \cr
  $  1$ & $A_2$ & $O_{1234}$ & $3$ & $ 9$-$16$ & $0.69$ & $0.5983(31)(37)$ & $0.5652(33)(39)$ & $6.1(7.1)(8.3)$ &  No  \cr
\hline
  $  1$ & $E $ & $O_{123 }$ & $1$ & $ 8$-$18$ & $1.43$ & $0.5004(18)(14)$ & $0.4603(20)(16)$  & $93.7(1.7)(1.3)$ &  Yes  \cr
  $  1$ & $E $ & $O_{123 }$ & $2$ & $ 8$-$17$ & $1.37$ & $0.6136(25)(24)$ & $0.5813(27)(26)$ & $166.3(2.8)(2.7)$ &  Yes  \cr
\hline
  $  \sqrt{2}$ & $B1$ & $O_{1234}$ & $1$ & $ 8$-$18$ & $1.23$ & $0.5041(13)(10)$ & $0.4207(16)(12)$  & $8.84(89)(68)$ &  Yes  \cr
  $  \sqrt{2}$ & $B1$ & $O_{1234}$ & $2$ & $ 8$-$17$ & $1.09$ & $0.5557(26)(27)$ & $0.4814(30)(31)$  & $144.9(2.3)(2.4)$ &  Yes  \cr
\hline
  $  \sqrt{2}$ & $B2$ & $O_{1234}$ & $1$ & $ 8$-$18$ & $0.56$ & $0.5189(15)(11)$ & $0.4384(18)(13)$ & $19.9(1.7)(1.2)$ &  Yes  \cr
  $  \sqrt{2}$ & $B2$ & $O_{1234}$ & $2$ & $ 8$-$18$ & $1.18$ & $0.5634(26)(23)$ & $0.4902(30)(27)$ & $152.0(2.6)(2.4)$ &  Yes  \cr
  $  \sqrt{2}$ & $B2$ & $O_{1234}$ & $3$ & $ 8$-$16$ & $1.28$ & $0.6717(40)(49)$ & $0.6116(44)(54)$ & $158(14)(17)$ &  No  \cr
\hline
  $  \sqrt{2}$ & $B3$ & $O_{1234}$ & $1$ & $ 9$-$18$ & $0.97$ & $0.5376(38)(34)$ & $0.4603(45)(39)$ & $99.1(3.5)(3.1)$ &  Yes  \cr
  $  \sqrt{2}$ & $B3$ & $O_{1234}$ & $2$ & $ 9$-$18$ & $1.15$ & $0.6573(43)(49)$ & $0.5958(48)(54)$ & $174(15)(172)$ &  No  \cr
  $  \sqrt{2}$ & $B3$ & $O_{1234}$ & $3$ & $ 8$-$14$ & $0.82$ & $0.6780(67)(88)$ & $0.6185(74)(96)$ & $167.0(5.6)(6.9)$ &  No  \cr
\hline
  $  \sqrt{3}$ & $A2$ & $O_{1234}$ & $1$ & $ 8$-$18$ & $0.68$ & $0.5538(35)(49)$ & $0.4371(44)(62)$ & $15.5(3.4)(4.8)$ &  Yes  \cr
  $  \sqrt{3}$ & $A2$ & $O_{1234}$ & $2$ & $ 8$-$16$ & $1.41$ & $0.5905(35)(39)$ & $0.4827(43)(48)$ & $149(11)(13)$ &  Yes  \cr
  $  \sqrt{3}$ & $A2$ & $O_{1234}$ & $3$ & $ 8$-$16$ & $1.10$ & $0.6093(49)(50)$ & $0.5055(59)(60)$ & $156.5(7.5)(14.4)$ &  No  \cr
\hline
  $  \sqrt{3}$ & $E $ & $O_{123 }$ & $1$ & $ 8$-$16$ & $0.71$ & $0.5641(37)(41)$ & $0.4501(47)(50)$ & $44.4(5.0)(5.3)$ &  Yes  \cr
  $  \sqrt{3}$ & $E $ & $O_{123 }$ & $2$ & $ 7$-$16$ & $0.72$ & $0.6195(33)(54)$ & $0.5178(39)(64)$ & $160.6(3.3)(5.4)$ &  Yes  \cr
\hline
\end{tabular}
\caption{\label{tab:gevp_results} GEVP results for the energy levels. We set $t_0/a=3$ and
use the one-exponential form in Eq.~(\ref{oneexp}) to fit the principal correlators. Also
shown are the corresponding center-of-mass energy $\sqrt{s_n^{\Lambda,\, \vec{P}}}$ and
extracted phase shift $\delta_1\Big(\sqrt{s_n^{\Lambda,\, \vec{P}}}\:\Big)$. The last
column indicates whether the energy level is used our global analysis of $\pi\pi$
scattering (see Sec.~\ref{sec_results}).}
\end{table*}

The generalized eigenvalue problem is defined as
\begin{align}
\label{eq:GEVP}
C^{\Lambda,\vec{P}}_{ij}(t)u^n_j(t) = \lambda^n(t,t_0) C^{\Lambda,\vec{P}}_{ij}(t_0)u_j^n(t),
\end{align}
where $t_0$ is a reference time
\cite{Michael:1985ne,Luscher:1990ck,Blossier:2009kd,Orginos:2015tha}. At large $t$, the
eigenvalues $\lambda^{n}(t,t_0)$, which are also referred to as principal correlators,
behave as
\begin{align}
\label{oneexp}
\lambda^{n}(t,t_{0}) = \E^{-E_{n}^{\Lambda,\vec{P}}(t-t_{0})}.
\end{align}
To determine the energies $E_n^{\Lambda,\vec{P}}$, we fit the eigenvalues either with the
single-exponential form of Eq.~\refeq{oneexp} or with the two-exponential form
\begin{align}
\label{twoexp}
\lambda^{n}(t,t_{0}) = (1-B) \E^{-E_{n}^{\Lambda,\vec{P}}(t-t_{0})} + B \E^{-E_{n}^{'\Lambda,\vec{P}}(t-t_{0})},
\end{align}
which perturbatively includes a small pollution from higher-lying excited states with
energies $E_n^{'\Lambda,\vec{P}}$ \cite{Luscher:1990ck,Blossier:2009kd}. We checked the
GEVP spectrum for $t_0/a \in [2,9]$ and found that the central values are independent of
$t_0$ within statistical uncertainties. We set $t_0/a=3$ for our main analysis, which
minimizes the overall statistical noise. The chosen fit types, fit ranges, corresponding
$\chi^2$ values, the energies, and other derived quantities are presented in Table
\ref{tab:gevp_results}. The operator basis used is $O_{1234}$ in all irreps except
$E$, where we only use $O_{123}$ because the energy level dominantly overlapping
with $O_4$ is too far above the region of interest.

For each quantity $y$, the first uncertainty given is the statistical uncertainty,
obtained from single-elimination jackknife. The second uncertainty is the systematic
uncertainty, estimated using the prescription
\begin{align}
\sigma^{sys}_y = {max} \bigg( |y_{avg}^{\prime} - y_{avg}|,\sqrt{|\sigma_y^{\prime 2} - \sigma_y^2|} \bigg) , \label{eq:sigmasys}
\end{align}
where $y_{avg}$ and $\sigma_y$ are the central value and statistical uncertainty for the
chosen fit range specificed in Table
\ref{tab:gevp_results}, and $y_{avg}^{\prime}$, $\sigma_y^\prime$ are the central value and statistical uncertainty obtained with $t_{min}/a$
increased by $1$.

\subsection{Matrix fit analysis}\label{subsec_MFA}

\begin{figure}[hb!]
\begin{center}
 \includegraphics[width=\linewidth]{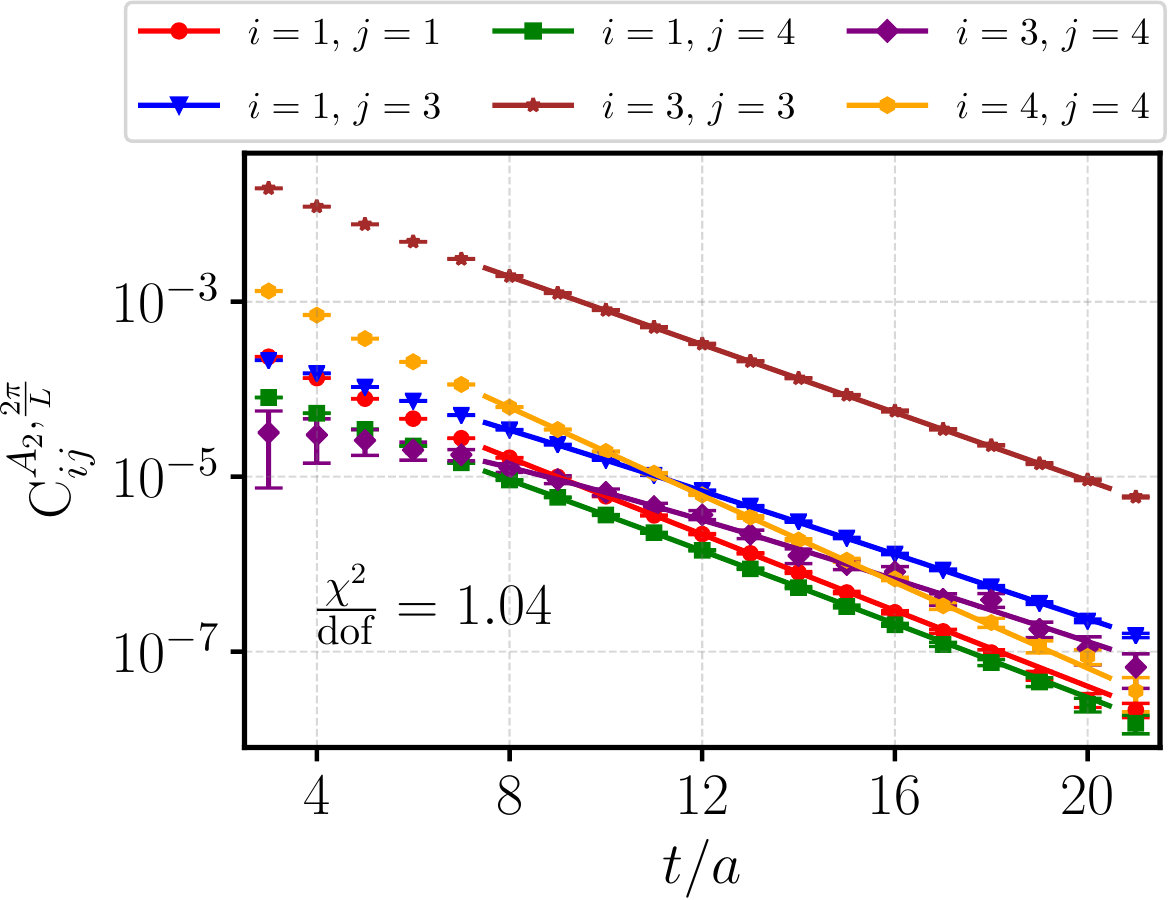}

\caption{\label{fig:A2MFA} Sample matrix fit with $N_{states}=3$ for
$|\vec{P}|=\frac{2\pi}{L},\Lambda=A_2$ in the range between $t_{min}/a=8$ and
$t_{min}/a=20$.}

\end{center}
\end{figure}

The spectral decomposition of the correlation matrix (neglecting the finite time extent of
the lattice) reads
\begin{equation}
 C^{\Lambda,\vec{P}}_{ij}(t) = \sum_{n=1}^{\infty} \langle 0 | O_i | n, \Lambda,\vec{P} \rangle\: \langle n, \Lambda,\vec{P} | O_j^\dag | 0 \rangle\: \E^{-E_n^{\Lambda,\vec{P}} t},
\end{equation}
where $| n, \Lambda,\vec{P} \rangle$ is the $n$-th energy eigenstate with the given
quantum numbers. We defined the interpolating fields $O_i$ such that the entire
correlation matrix $C^{\Lambda,\vec{P}}_{ij}(t)$ is real-valued (in the infinite-
statistics limit); this is possible because of charge-conjugation symmetry. Consequently,
the overlap factors $Z_{i,\,n}=\langle 0 | O_i | n, \Lambda,\vec{P} \rangle$ can also be
chosen as real-valued. In the matrix fit analysis, we directly fit the correlation matrix
for $t_{min} \leq t \leq t_{max}$ using the model
\begin{eqnarray}
 C^{\Lambda,\vec{P}}_{ij}(t) &\approx & \sum_{n=1}^{N_{states}} Z_{i,\,n}\: Z_{j,\,n}\: \E^{-E_n^{\Lambda,\vec{P}} t},
\end{eqnarray}
where $t_{min}$ has to be chosen large enough such that contributions from $n>N_{states}$
become negligible. For an $m\times m$ correlation matrix, this model has $N_{states}\times
(m+1)$ parameters. To ensure that the energies returned from the fit are ordered, we used
the logarithms of the energy differences, $l_{n}^{\Lambda,\vec{P}} = \ln\left(a
E_{n}^{\Lambda,\vec{P}}-aE_{n-1}^{\Lambda,\vec{P}}\right)$, instead of $a
E_n^{\Lambda,\vec{P}}$ (for $n>1$) as parameters in the fit. To simplify the task of
finding suitable start values for the iterative $\chi^2$-minimization process, we also
rewrote the overlap parameters as $Z_{i,\,n}=B_{i,\,n}Z_{i}$ with $B_{i,\,n}=1$ for $n$
equal to the state with which $O_i$ has the largest overlap. Good initial guesses for
$Z_i$ can then be obtained from single-exponential fits of the form $Z_i Z_i
\E^{-E_n^{\Lambda,\vec{P}} t}$ to the diagonal elements $C^{\Lambda,\vec{P}}_{ii}(t)$ in
an intermediate time window in which the $n$-th state dominates, and the start values of
$B_{i,\,n}$ can be set to zero. An example matrix fit is shown in Fig.~\ref{fig:A2MFA}.

In the matrix fits, we excluded the interpolating fields $O_2$, which are very similar to
$O_1$ and did not provide useful additional information. For each $(\Lambda,\vec{P})$, we
performed either $3\times3$ matrix fits (including $O_1$, $O_3$, $O_4$) with
$N_{states}=3$ or $2\times2$ matrix fits (including $O_1$ and $O_3$) with $N_{states}=2$.
We set $t_{max}=20$ and varied $t_{min}.$ The matrix fit results for $a
E_n^{\Lambda,\vec{P}}$ are shown as the black diamonds in the right panels of
Figs.~\ref{fig:SpecCompare} and \ref{fig:SpecCompare2}. We observe that the results for
all extracted energy levels stabilize for $t_{min}\gtrsim 8$.

\subsection{Comparison between GEVP and MFA}\label{subsec_compare}

\begin{figure*}
  \includegraphics[scale=0.38]{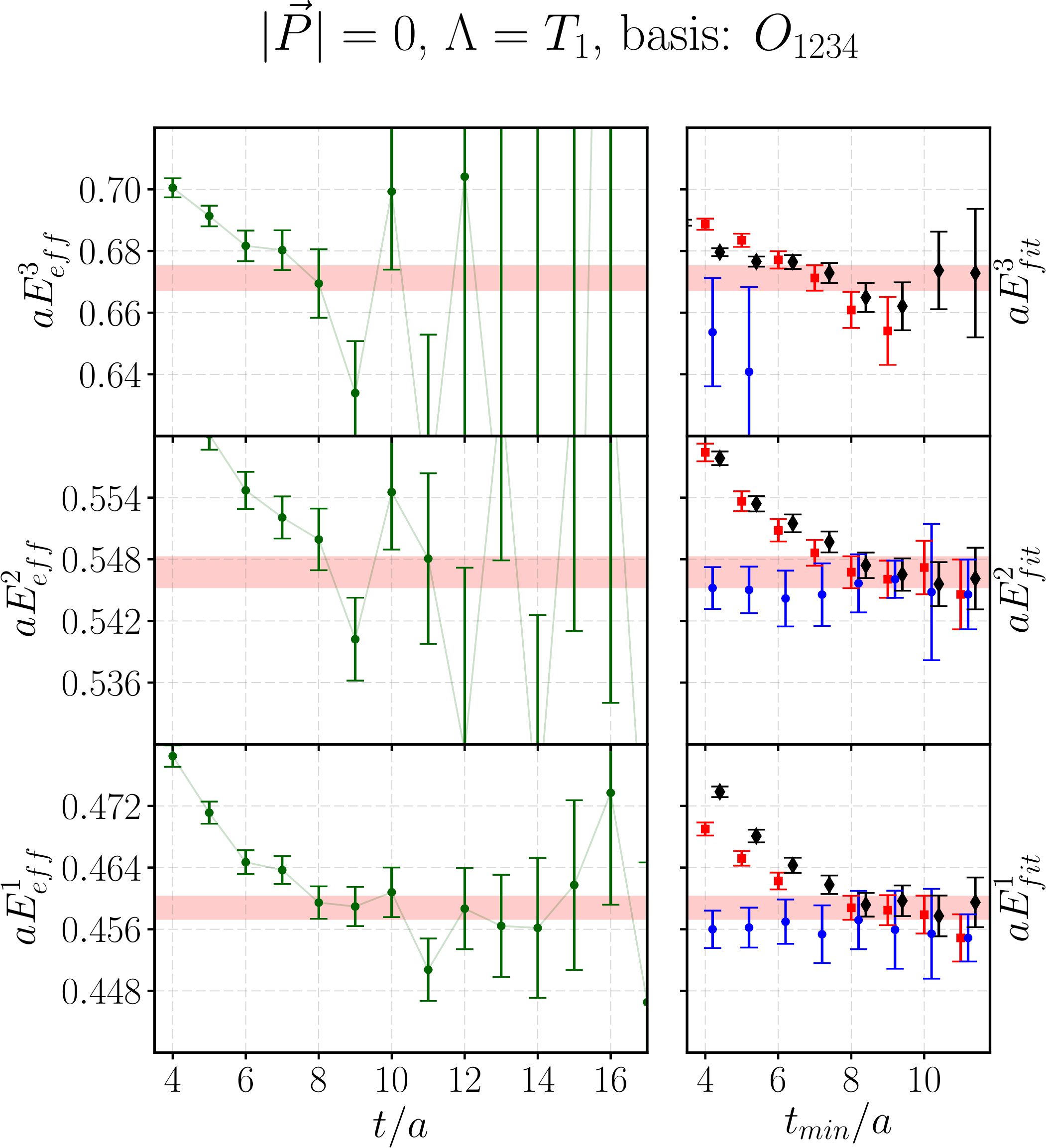} \hfill
  \includegraphics[scale=0.38]{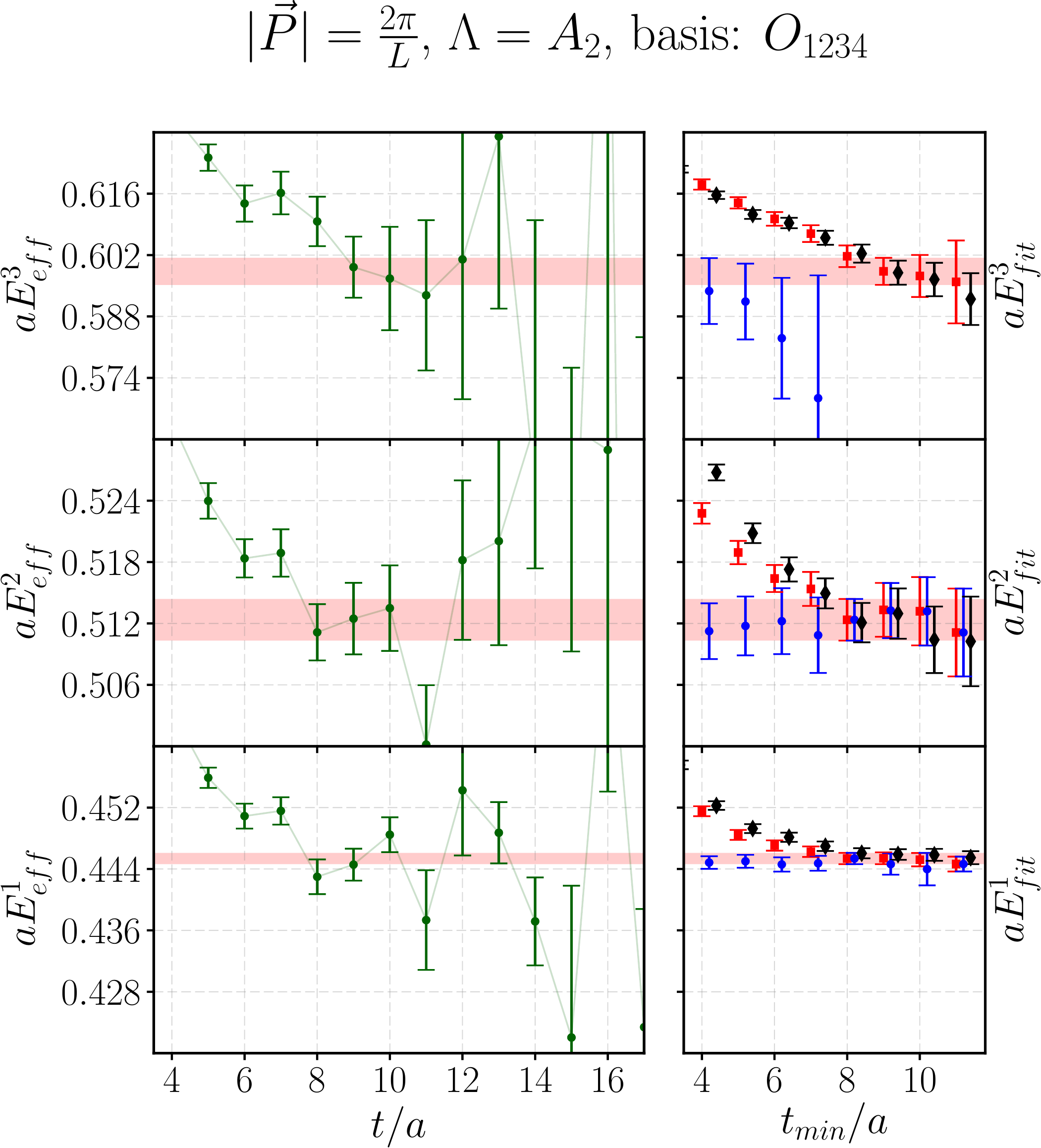}

  \includegraphics[scale=0.38]{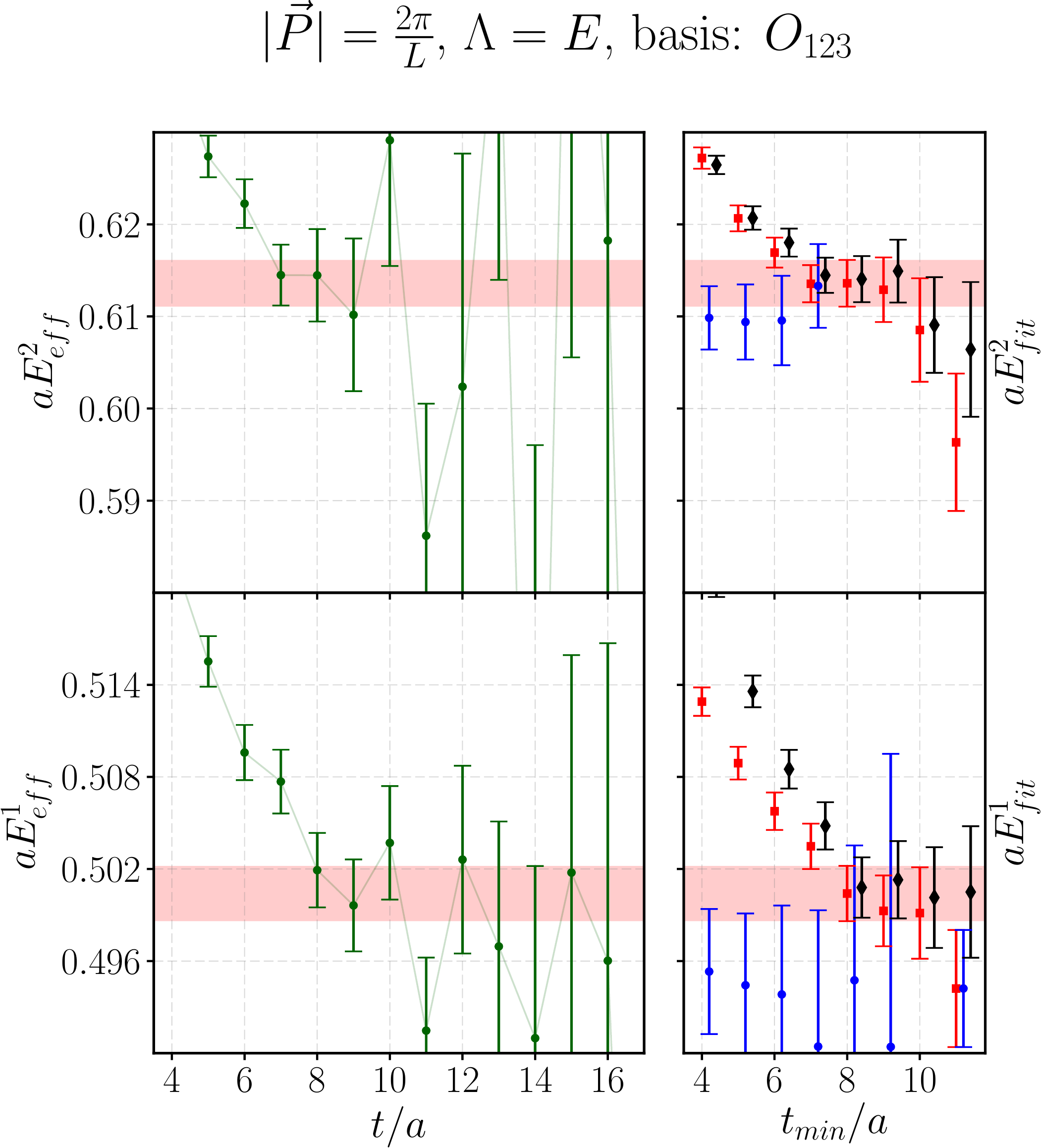} \hfill
  \includegraphics[scale=0.38]{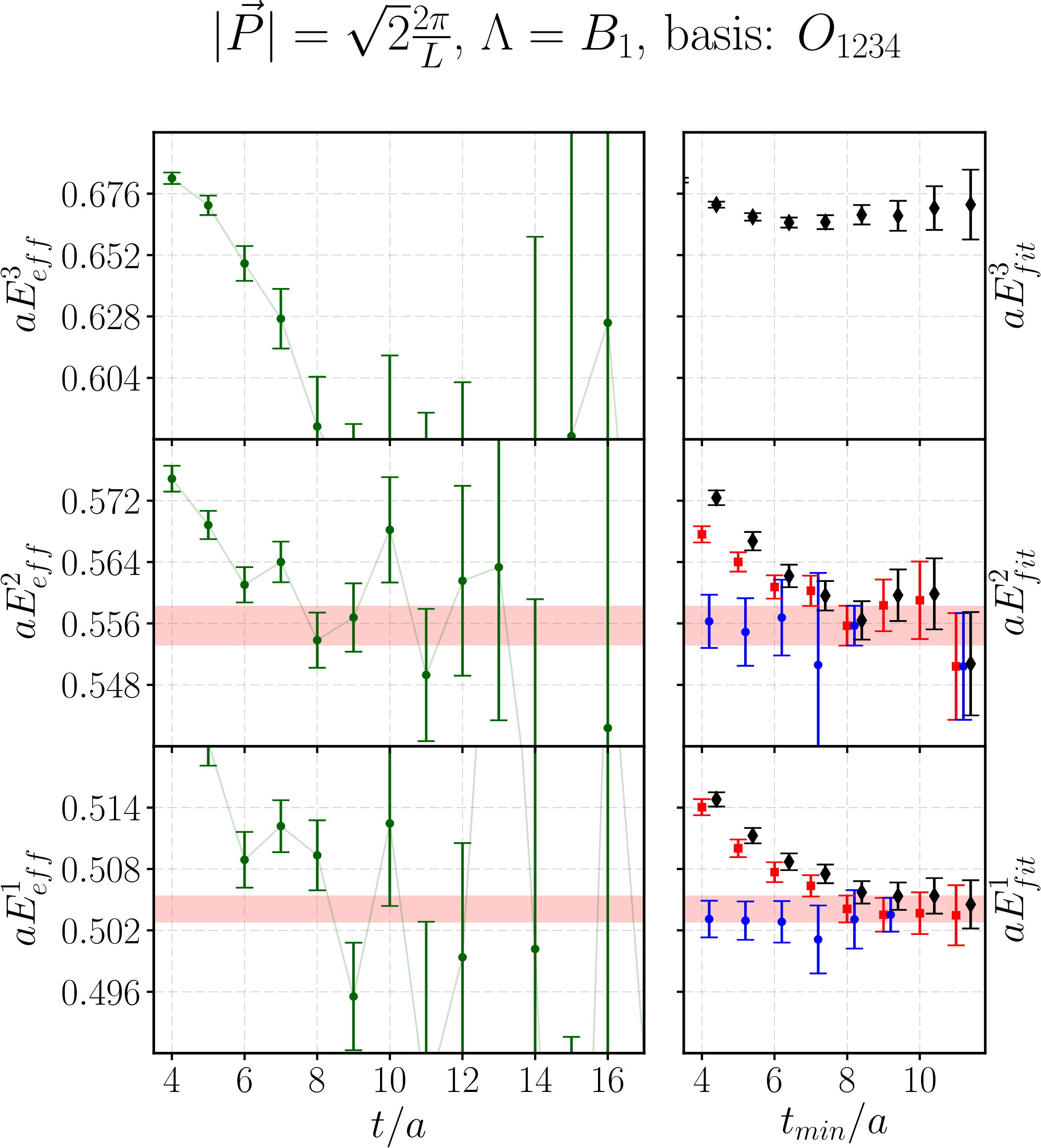}
  \caption{\label{fig:SpecCompare}Comparison between MFA and GEVP for the momentum frames
  and irreps $\frac{L}{2\pi}|\vec{P}|=0,1,\sqrt{2}$ and $\Lambda=T_1,A_2,E,B_1$,
  respectively. The green circles on the left panel show the effective energies
  $E_{eff}^{n}$ determined from the principal correlators. In the right panel we present
  the fitted energies as they depend on the choice of $t_{min}$. Black diamonds are
  obtained from MFA, red squares are obtained from the single exponential fits to the
  principal correlator [see Eq.~(\ref{oneexp})], and blue circles are from two-exponential
  fits to the principal correlator [see Eq.~(\ref{twoexp})]. Note that not all two-exponential
  fits are shown, as they can become unstable. The red horizontal bands give the $1\sigma$
  statistical-uncertainty ranges of the selected one-exponential GEVP fits listed in
  Table \protect\ref{tab:gevp_results}.}
\end{figure*}

\begin{figure*}
  \includegraphics[scale=0.38]{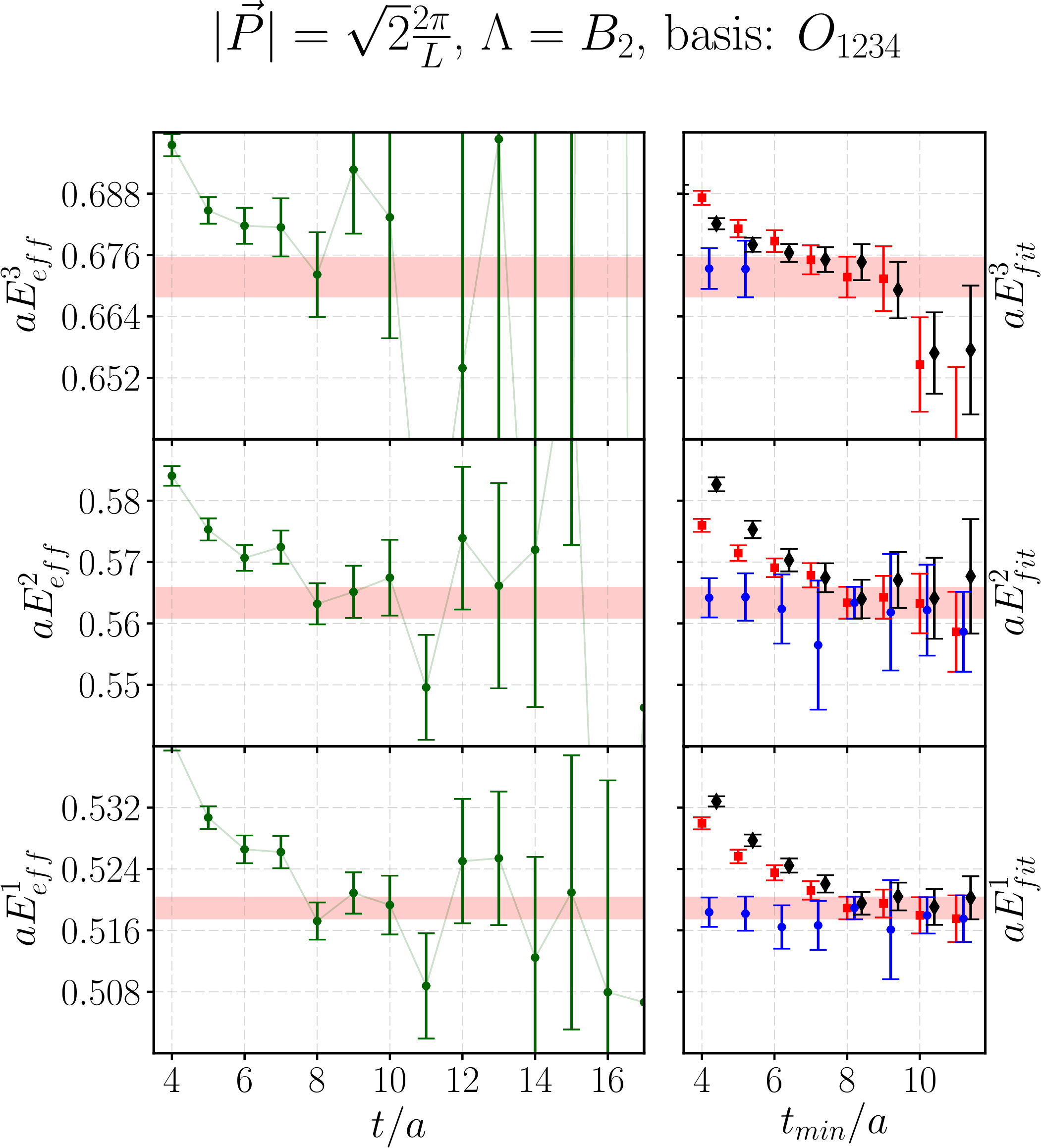} \hfill
  \includegraphics[scale=0.38]{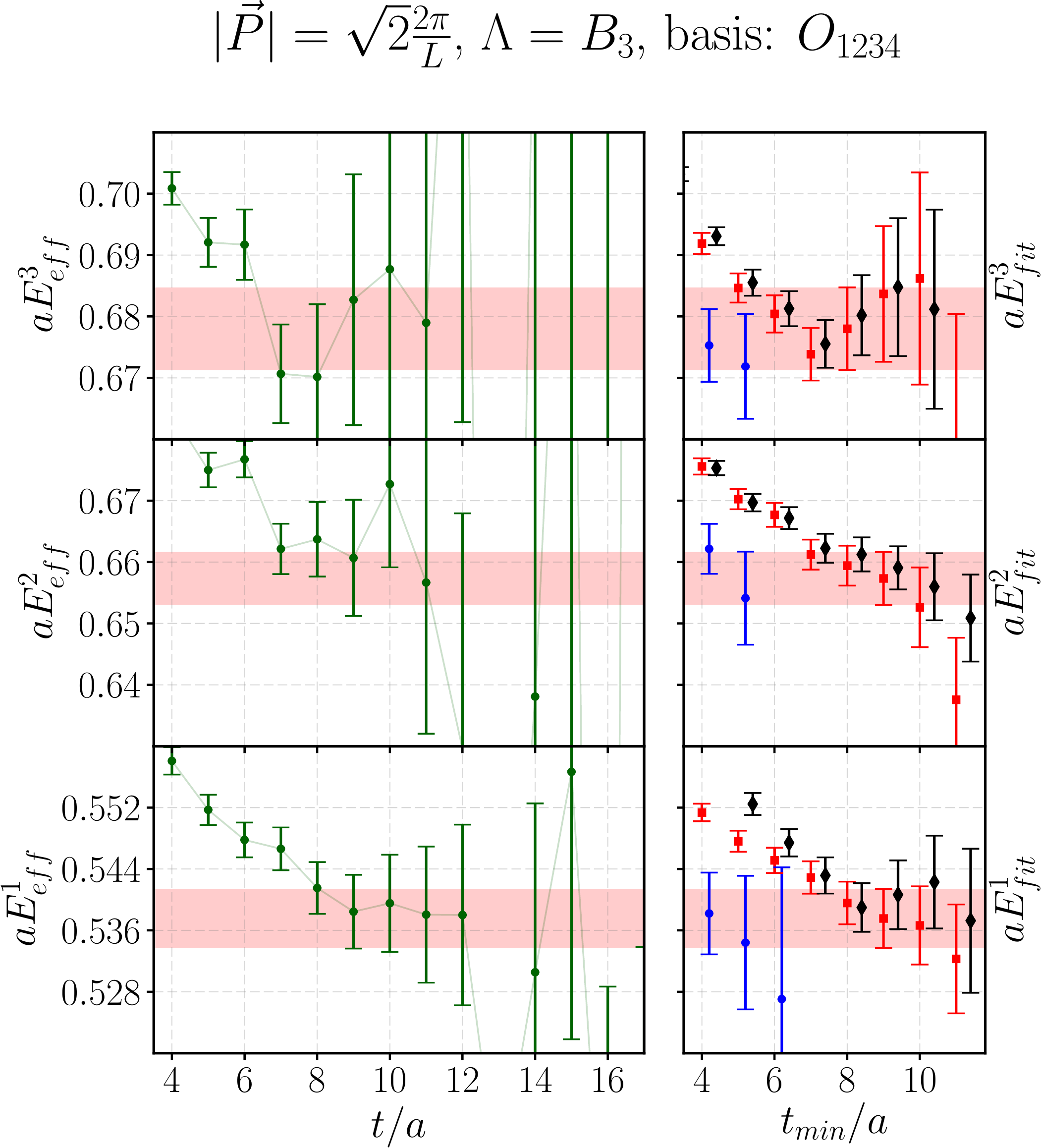}

  \includegraphics[scale=0.38]{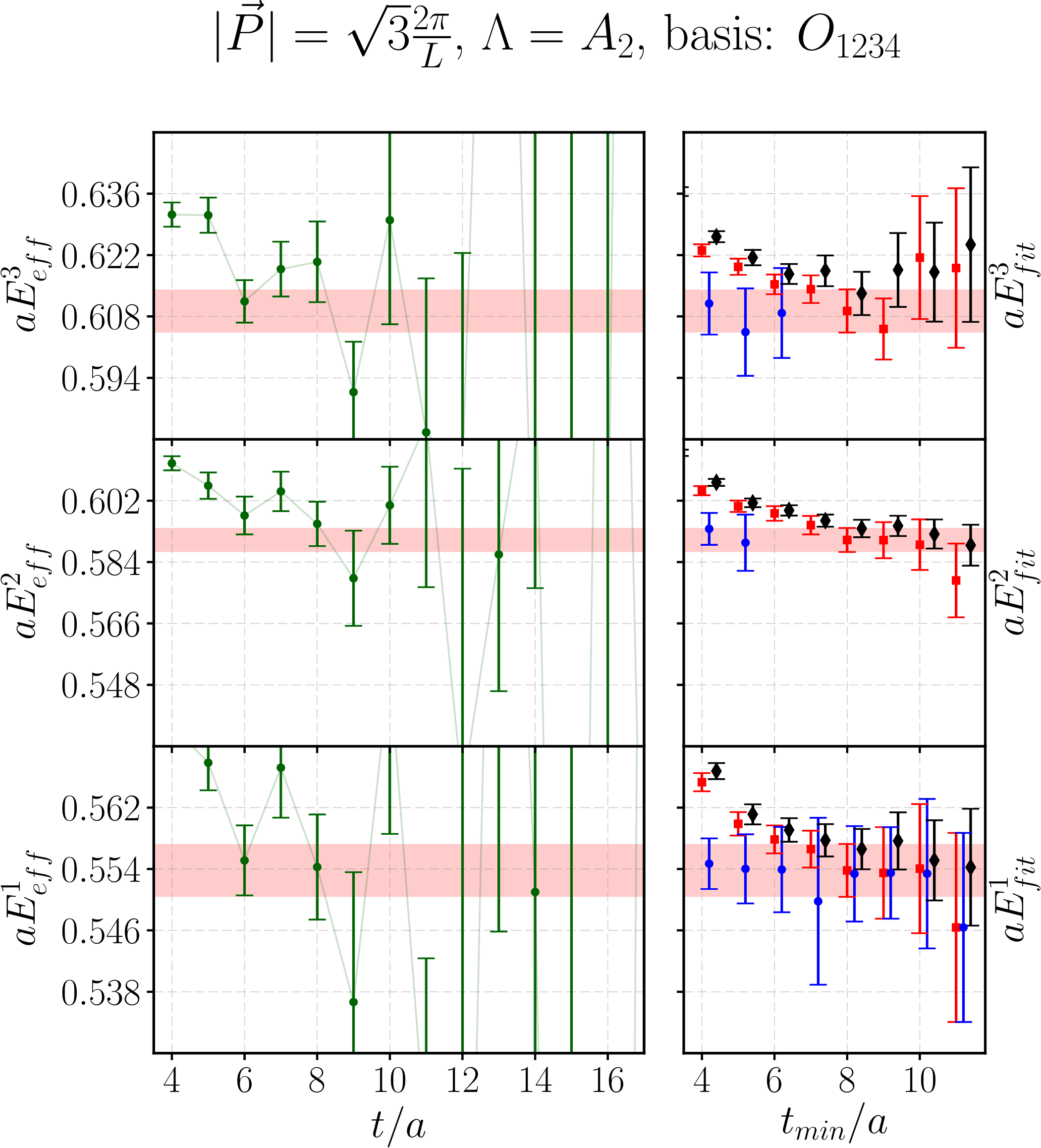} \hfill
  \includegraphics[scale=0.38]{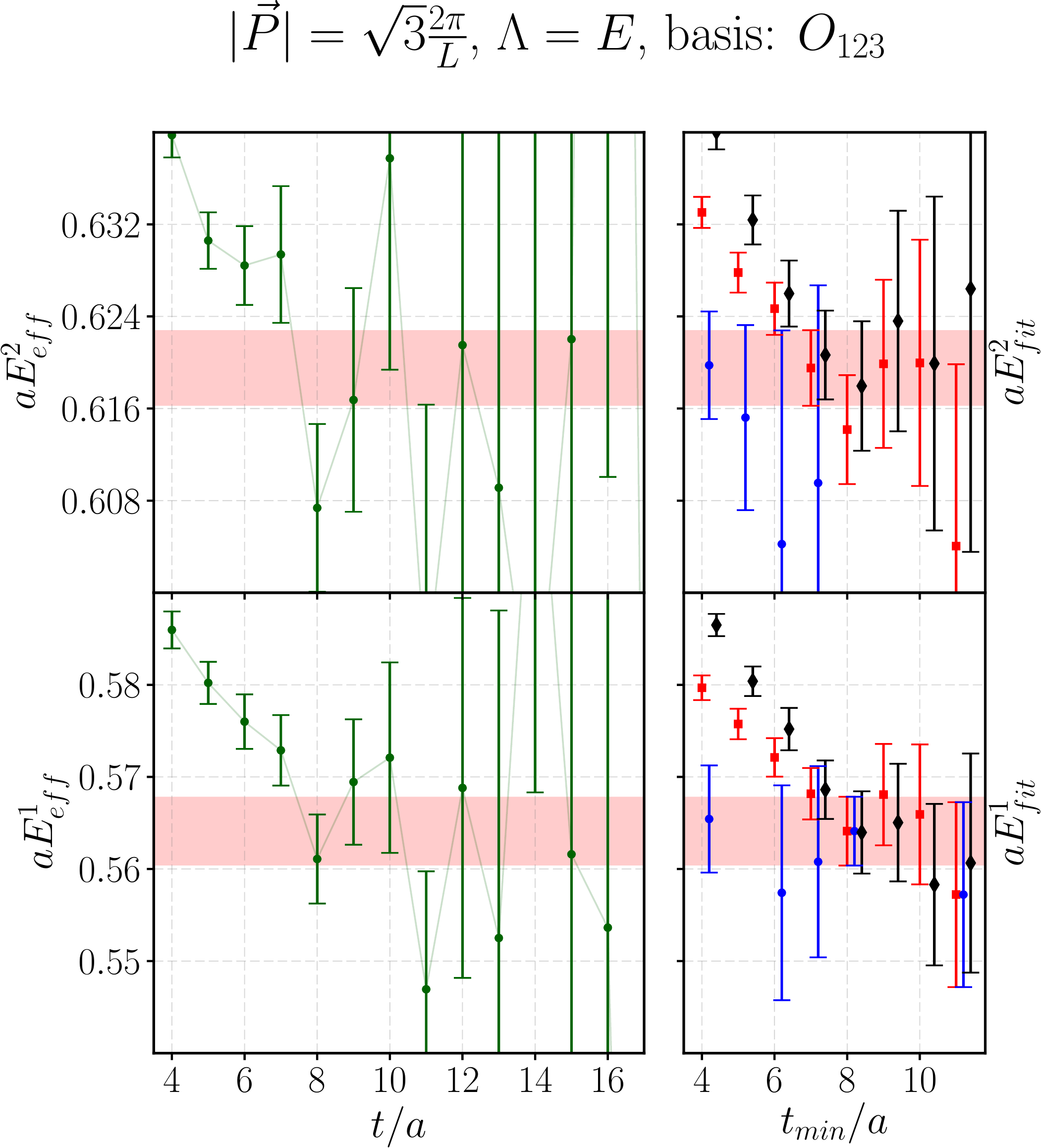}
  \caption{\label{fig:SpecCompare2}As in Fig.~\protect\ref{fig:SpecCompare}, but for
  $\frac{L}{2\pi}|\vec{P}|=\sqrt{2},\sqrt{3}$ and $\Lambda=B_2,B_3,A_2,E$.}
\end{figure*}

\begin{figure*}

  \null

  \vspace{-8ex}

  \includegraphics[scale=0.37]{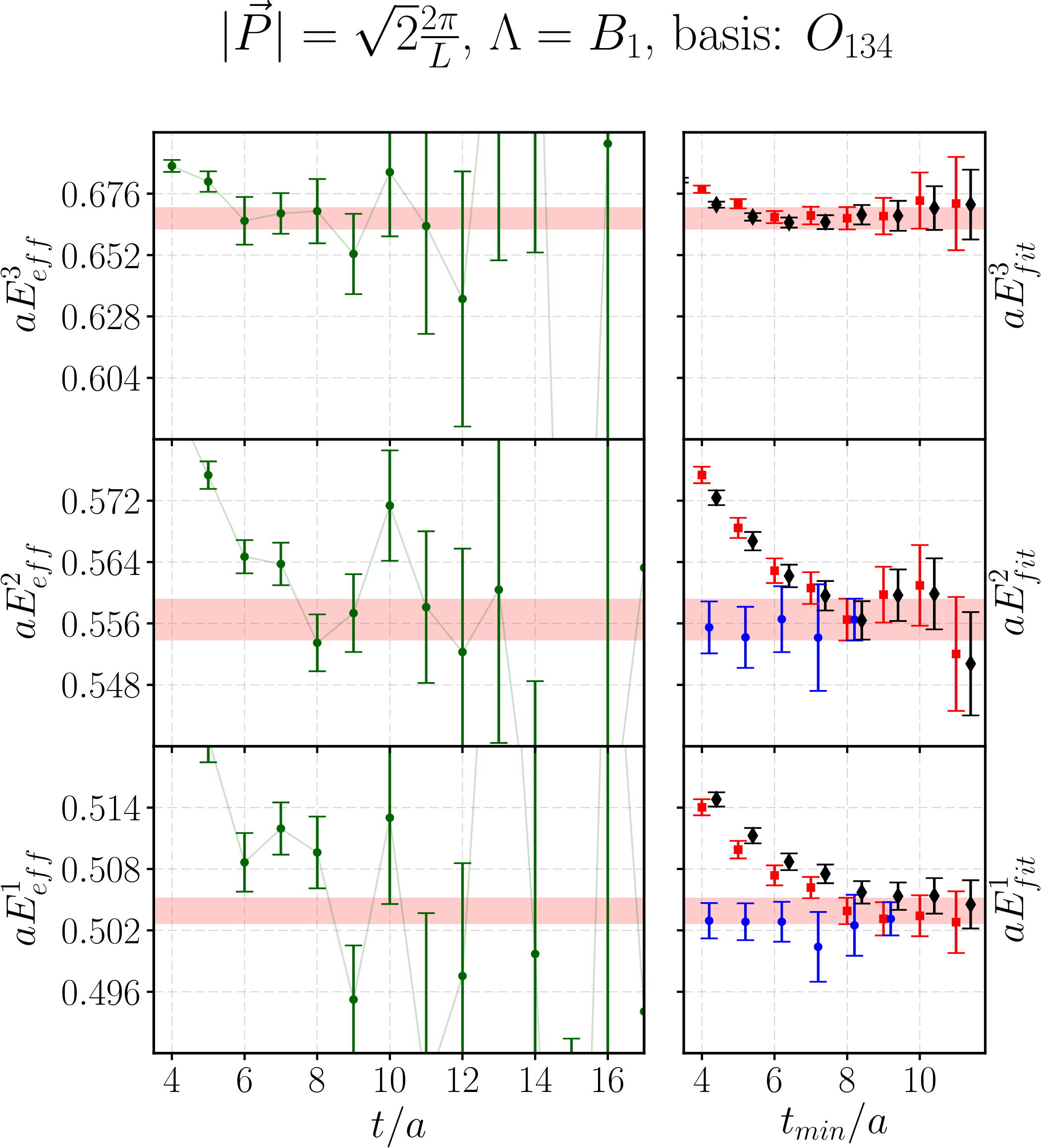}

  \caption{\label{fig:B1lowbase} Comparison between MFA and GEVP for the $B_1$ irrep with
  $|\vec{P}|=\sqrt{2}\frac{2\pi}{L}$ as in Fig.~\protect\ref{fig:SpecCompare}, but with
  $O_2$ removed from the basis for the GEVP. The reduced basis gives a better extraction
  of $a E_3$ compared to the full basis only in this irrep. }
\end{figure*}

The results obtained from the GEVP and the MFA are compared in Figs. \ref{fig:SpecCompare}
and \ref{fig:SpecCompare2}. The left panels show the effective energy
\begin{equation}
aE_{eff}^{n}(t) = \ln \frac{\lambda_n(t,t_0)}{\lambda_n(t+a,t_0)}
\end{equation}
of the GEVP principal correlators, while the right panes show the fit results $aE^n_{fit}$
from both the GEVP and the MFA as a function of $t_{min}$ (we did not find any significant
dependence on $t_{max}$). For the GEVP, we show both one- and two-exponential fits using
Eqs.~(\ref{oneexp}) and (\ref{twoexp}). We find that the one-exponential GEVP fit results
are very similar (both in central value and uncertainty) to the MFA results, except for
the $n=3$ energy level of the $|\vec{P}|=\sqrt{2}\frac{2\pi}{L}, \Lambda=B_1$ correlation
matrix where the principal correlator obtained from the GEVP with the basis $O_{1234}$
does not show a plateau and we do not extract this energy level. Surprisingly, we found
that removing the second quark-antiquark operator $O_2$ from the basis yields a stable
plateau and stable fit results for the $n=3$ energy level, as shown in
Fig.~\ref{fig:B1lowbase}. Note that $O_2 \sim \bar{q}\gamma_0\gamma_i q$ has a very
similar structure as $O_1 \sim \bar{q}\gamma_i q$. For $n=1$ and $n=2$, the
one-exponential fit results for the chosen
$t_{\rm min}/a=8$ change by less than $0.5\sigma$
when removing $O_2$. We also performed additional GEVP fits with the reduced basis in all
other irreps, and found that none of the fitted energies changed significantly (in fact,
the reduced basis gives slightly larger uncertainties in most cases). Given that the $n=3$
energy in the $B_1$ irrep is above the $4\pi$ and $K\bar{K}$ thresholds, we do not use
this energy level in our further analysis.

Finally, we note that the two-exponential fits to the GEVP principal correlators find
plateaus at much smaller $t_{min}$ but are significantly noisier compared to the MFA and
one-exponential GEVP fits. Overall, we have shown that the MFA and GEVP methods are
equivalent, and we use the one-exponential GEVP fit results given in Table
\ref{tab:gevp_results} in our further analysis. These results are also indicated with the
red bands in Figs. \ref{fig:SpecCompare} and \ref{fig:SpecCompare2}.

\section{The  L\"uscher analysis: formalism}\label{sec_formalism}

Even though we have some energy levels with quite large invariant mass (see Table
\ref{tab:gevp_results}), we limit our energy region of interest below $0.55 a^{-1}$ where we are
safely away from the $4\pi$ ($\approx 0.73$) and $K\bar{K}$ ($\approx 0.6$) thresholds
\cite{Chen:2012rp} and can safely perform the elastic scattering analysis of the L\"uscher method.

The quantization condition for elastic $\pi\pi$ scattering is
\begin{align}
\label{eq:QC}
{\rm det} \bigg( \mathbbm{1} + \I t_{\ell}(s)(\mathbbm{1} + \I {\cal M}^{\vec{P}}) \bigg) = 0,
\end{align}
where $t_{\ell}(s)$ is the infinite-volume scattering amplitude, which is related to the
infinite-volume scattering phase shift $\delta_{\ell}(s)$ via Eq.~(\ref{eq:tlofs}). The
matrix ${\cal M}^{\vec{P}}$ has the indices ${\cal M}_{lm,l'm'}^{\vec{P}}$, where $l,l'$
label the irreducible representations of $SO(3)$ and $m,m'$ are the corresponding row
indices. For the case of $P$-wave $\pi\pi$ scattering, $F$-wave and higher contributions
are highly suppressed, as was shown in a previous lattice study \cite{Dudek:2012xn} and in
an analysis of experimental data \cite{Estabrooks:1975cy}. Neglecting these contributions,
the matrix ${\cal M}^{\vec{P}}$ takes the form
\begin{widetext}
\begin{align}
\label{eq:Mmat}
{\cal M}^{\vec{P}} = \bordermatrix{~ & 0\,0        & 1\,0                    & 1\,1                       & 1\,-\!1\cr
                            0\:0 & w_{00} & i\sqrt{3}w_{10} & i \sqrt{3} w_{11} & i\sqrt{3} w_{1-1} \cr
                            1\:0 & -i\sqrt{3}w_{10} & w_{00}+2w_{20} & \sqrt{3} w_{21} & \sqrt{3}w_{2-1} \cr
                            1\:1 & i \sqrt{3}w_{1-1} & -\sqrt{3}w_{2-1} & w_{00}-w_{20} & -\sqrt{6} w_{2-2} \cr
                            1\,-\!1 & i \sqrt{3} w_{11} & -\sqrt{3}w_{21} & -\sqrt{6}w_{22} & w_{00}-w_{20} \cr},\
\end{align}
\end{widetext}
where the indices $lm$ and $l'm'$ are indicated next to the matrix. The functions $w_{lm}$
are equal to
\begin{align}
\label{eq:defw}
w_{lm}^{\vec{P}}(k,L) = \frac{ Z_{lm}^{\vec{P}}\left(1;(k L/(2 \pi))^2\right) }{ \pi^{3/2} \sqrt{2l+1} \gamma (\frac{k L}{2 \pi})^{l+1}},
\end{align}
where $Z_{lm}^{\vec{P}}(1;(\frac{k L}{2 \pi})^2)$ is the generalized zeta function as
defined for example in Appendix A of Ref.~\cite{Leskovec:2012gb}, and $\gamma=E/\sqrt{s}$
is the Lorentz boost factor. The matrix ${\cal M}^{\vec{P}}$ can be further simplified by
taking into account the symmetries for a given Little Group ($\vec{P}$) and its irrep
$\Lambda$ \cite{Leskovec:2012gb}. The quantization condition (\ref{eq:QC}) then reduces to
the following equations for each $\vec{P}$ and $\Lambda$:

\begin{widetext}
\small
\begin{align}
\nonumber &\vec{P} = 0, \;\;\,\; \Lambda = T_1 \text{: }\cr
&\; \cot{\delta_1(s_n^{\Lambda, \vec{P}})} =  w_{0,0}\left( \qqvec \right) \\
&\vec{P} = \frac{2\pi}{L} (0,0,1), \; \Lambda = A_2 \text{: }\cr
&\; \cot{\delta_1(s_n^{\Lambda, \vec{P}})} =  w_{0,0}\left( \qqvec \right) + 2 w_{2,0}\left( \qqvec \right) \cr
&\vec{P} = \frac{2\pi}{L} (0,0,1), \; \Lambda = E \text{: }\cr
&\; \cot{\delta_1(s_n^{\Lambda, \vec{P}})} =  w_{0,0}\left( \qqvec \right) -  w_{2,0}\left( \qqvec \right) \cr
&\vec{P} = \frac{2\pi}{L} (0,1,1), \; \Lambda = B_1 \text{: }\cr
&\; \cot{\delta_1(s_n^{\Lambda, \vec{P}})} =  w_{0,0}\left( \qqvec \right) + \frac{1}{2} w_{2,0}\left( \qqvec \right) + \I \sqrt{6} w_{2,1}\left( \qqvec \right) - \sqrt{\frac{3}{2}} w_{2,2}\left( \qqvec \right) \cr
&\vec{P} = \frac{2\pi}{L} (0,1,1), \; \Lambda = B_2 \text{: }\cr
&\; \cot{\delta_1(s_n^{\Lambda, \vec{P}})} =  w_{0,0}\left( \qqvec \right) + \frac{1}{2} w_{2,0}\left( \qqvec \right) - \I \sqrt{6} w_{2,1}\left( \qqvec \right) - \sqrt{\frac{3}{2}} w_{2,2}\left( \qqvec \right) \cr
&\vec{P} = \frac{2\pi}{L} (0,1,1), \; \Lambda = B_3 \text{: }\cr
&\; \cot{\delta_1(s_n^{\Lambda, \vec{P}})} =  w_{0,0}\left( \qqvec \right) -  w_{2,0}\left( \qqvec \right) + \sqrt{6} w_{2,2}\left( \qqvec \right) \nonumber \cr
&\vec{P} = \frac{2\pi}{L} (1,1,1), \; \Lambda = A_2 \text{: }\cr
&\; \cot{\delta_1(s_n^{\Lambda, \vec{P}})} =  w_{0,0}\left( \qqvec \right) - \I \sqrt{\frac{8}{3}} w_{2,2}\left( \qqvec \right) -  \sqrt{\frac{8}{3}} \Bigg( {\rm Re}\left[ w_{2,1}\left( \qqvec \right)\right] + {\rm Im}\left[ w_{2,1}\left( \qqvec \right)\right] \Bigg) \cr
&\vec{P} = \frac{2\pi}{L} (1,1,1), \; \Lambda = E  \text{: } \\
&\; \cot{\delta_1(s_n^{\Lambda, \vec{P}})} =  w_{0,0}\left( \qqvec \right) + \I \sqrt{6} w_{2,2} \left( \qqvec \right).  \label{eq:phase_mapping}
\end{align}
\normalsize
\end{widetext}

The scattering analysis can be performed in two different ways, and in this work we
present a comparison between the methods:

\begin{itemize}
\item In the first approach, Eqs.~(\ref{eq:phase_mapping}) are used to map each
individual energy level ($s_n^{\Lambda, \vec{P}}$) to the corresponding value of the
scattering phase shift $\delta_1(s_n^{\Lambda, \vec{P}})$. One then fits a phase-shift
model to the extracted values of $\delta_1(s_n^{\Lambda, \vec{P}})$.

\item In the second approach, a model for the $t$-matrix is fitted directly to the
spectrum via the quantization condition \cite{Guo:2012hv}. This method has proven to be
quite successful in recent years \cite{Dudek:2012gj,Dudek:2012xn,Dudek:2014qha,Wilson:2014cna,Wilson:2015dqa,Dudek:2016cru,Briceno:2016mjc}.
Unlike the first approach, the $t$-matrix fit method is also well-suited for more
complicated coupled-channel analyses.

\end{itemize}

\section{The  L\"uscher analysis: results}\label{sec_results}

\subsection{Direct fits to the phases}\label{subsec_directfit}

\begin{figure}
  \includegraphics[width=0.99\columnwidth]{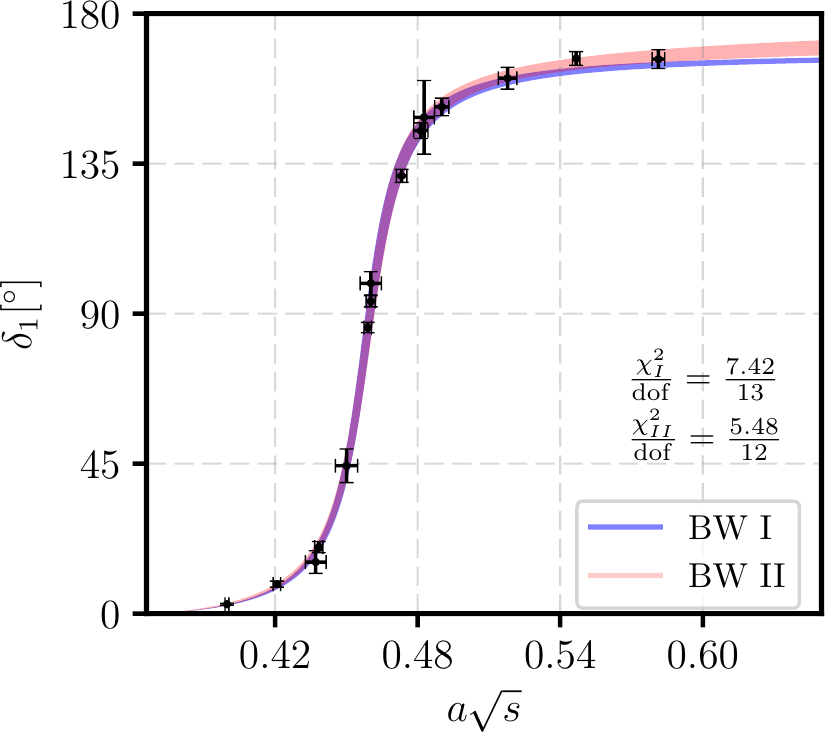}
  \caption{\label{fig:MIvMII}Comparison of fitting Breit-Wigner model {\bf BW I} versus
  fitting Breit-Wigner model {\bf BW II} to the phase shift data. The bands indicate the
  $1\sigma$ statistical uncertainty.}
\end{figure}

\begin{center}
\begin{table}
\begin{tabular}{| l | l | l | l | l | }
\hline
 Model & $\frac{\chi^2}{{\rm dof}}$ &  $am_\rho$ & $g_{\rho\pi\pi}$ & $(ar_0)^2$ \cr
\hline
{\bf BW I} & $0.571$ & $ 0.4599(19)(13) $  &  $ 5.76(16)(12) $ &   \cr
{\bf BW II} & $0.457$  &  $ 0.4600(18)(13) $  &  $ 5.79(16)(12) $  &  $ 8.6(8.0)(1.2) $ \cr
\hline
\end{tabular}
\caption{\label{tab:d_mIvsmII}Comparison of the parameters for the resonant Breit-Wigner models
 I and II.}
\end{table}
\end{center}

\begin{figure*}
  \includegraphics[width=0.95\textwidth]{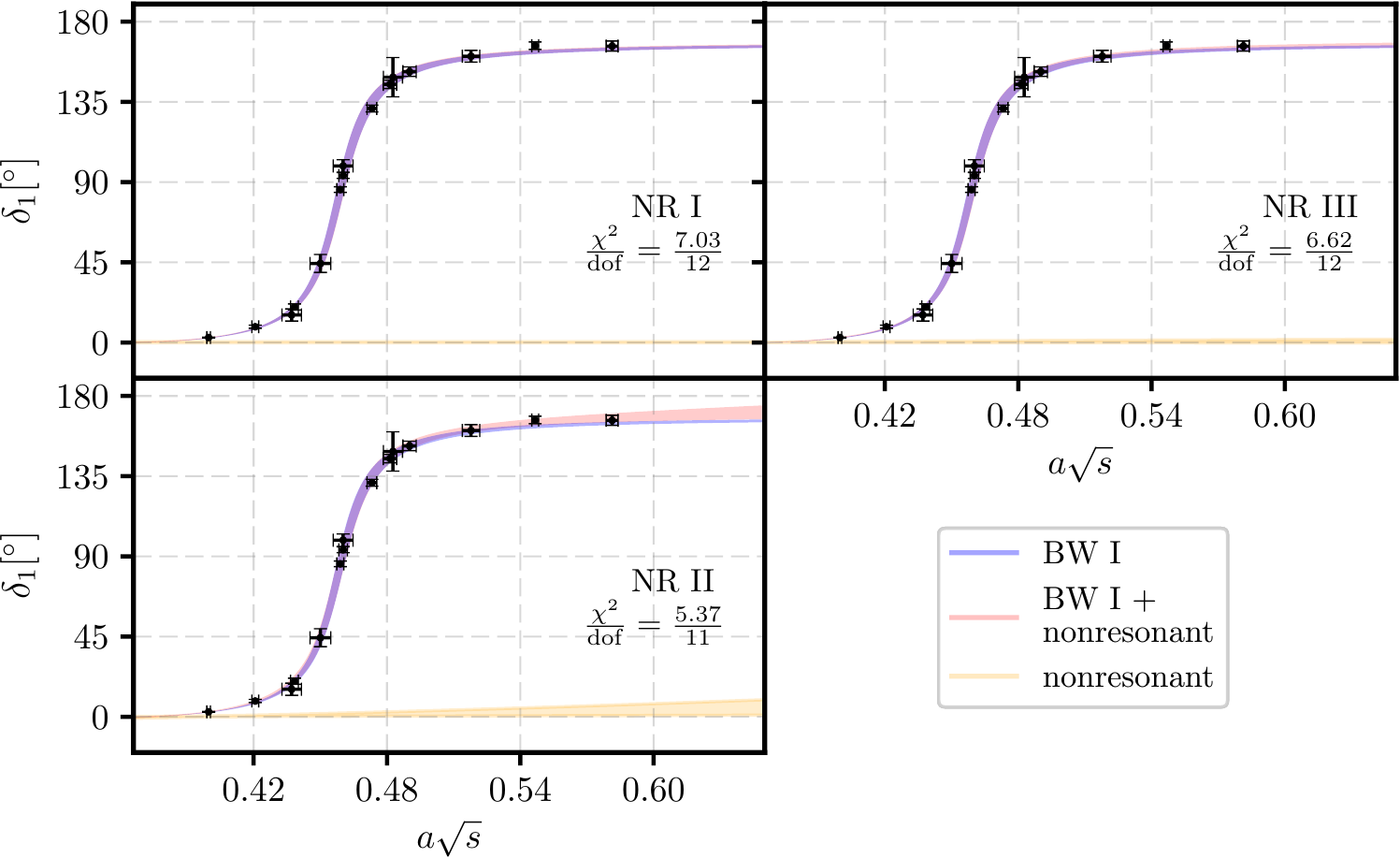}
  \caption{\label{fig:delta_MI} Contribution of nonresonant background models as
  described in Section \ref{sec_about_rho} to the resonant Breit-Wigner {\bf BW I}. None of the
  background phase shift models shows a strong sign of deviation away from $0$.}
\end{figure*}

\begin{table*}
\begin{tabular}{|l | l | l | l | l | l|}
\hline
 Model & $\frac{\chi^2}{{\rm dof}}$ &  $am_\rho$ & $g_{\rho\pi\pi}$  & & \cr
\hline
{\bf NR I } & $0.586$ &  $ 0.4600(19)(13) $  &  $ 5.74(17)(14) $  &  $ A  =  0.16(31)(18) ^\circ $ & \cr
{\bf NR II } & $0.488$ &  $ 0.4602(19)(13) $  &  $ 5.84(21)(20) $  &  $ A  =  -2.9(2.7)(3.4) ^\circ $  &  $ a^{-2}B =  19.2(16.6)(20.1)^\circ $ \cr
{\bf NR III } & $0.552$ &  $ 0.4601(19)(13) $  &  $ 5.74(16)(13) $  &  $ aa_1^{-1}  =  -19.8(27.4)(98.1) $ & \cr
\hline
\end{tabular}
\caption{\label{tab:mI_bg}Parameters of the phase shift model combining the resonant
Breit-Wigner model {\bf BW I} and various nonresonant models.}
\end{table*}

\begin{figure*}
  \includegraphics[width=0.95\textwidth]{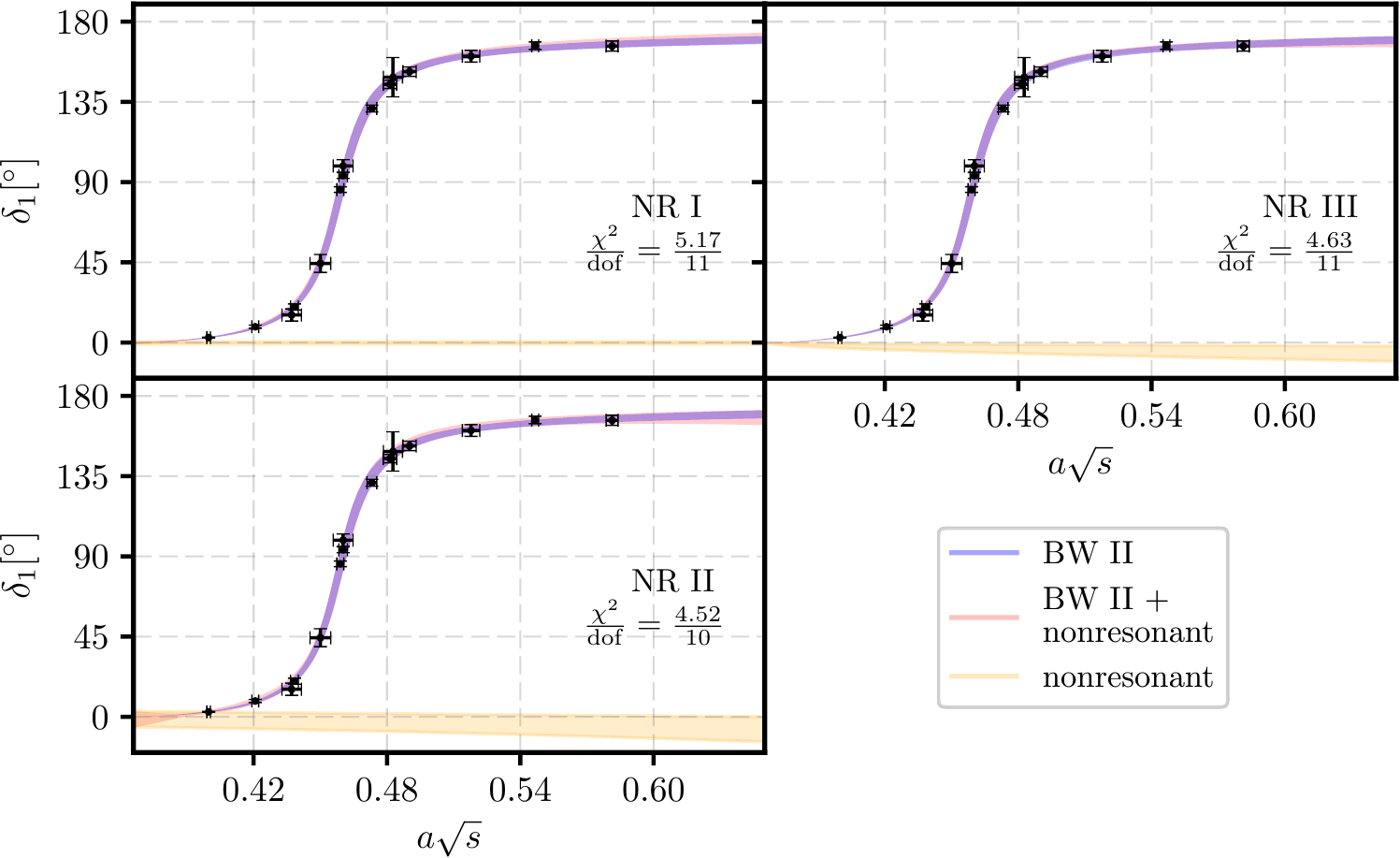}
  \caption{\label{fig:delta_MII} Contribution of nonresonant background models as
  described in Section \ref{sec_about_rho} to the resonant Breit-Wigner model {\bf BW II}. None of
  the background phase shift models shows a strong sign of deviation away from $0$.}
\end{figure*}

\begin{table*}
\resizebox{\textwidth}{!}{
\begin{tabular}{|l | l | l | l | l | l | l|}
\hline
 Model & $\frac{\chi^2}{{\rm dof}}$ & $am_\rho$ & $g_{\rho\pi\pi}$ & $(ar_0)^2$  & &\cr
\hline
{\bf NR I } & $0.470$ &  $ 0.4599(19)(26) $  &  $ 5.83(20)(21) $  &  $15.8(23.5)(1825.8) $  &  $A = -0.28(0.73)(12.56)^\circ $ &  \cr
{\bf NR II } & $0.452$ &  $ 0.4596(20)(14) $  &  $ 5.77(21)(20) $  &  $107.0(440.9)(631.0) $  &  $A = 1.3(4.5)(5.3) ^\circ$  &  $a^{-2}B = -19.8(16.0)(17.0) ^\circ$ \cr
{\bf NR III } & $0.421$ &  $ 0.4595(18)(8) $  &  $ 5.78(20)(9) $  &  $ 109.7(128.7)(117.6) $  &  $aa_1^{-1} = 2.4(1.7)(2.4) $ &  \cr
\hline
\end{tabular}
}
\caption{\label{tab:mII_bg}Parameters of the phase shift model combining the resonant
Breit-Wigner model {\bf BW II} and various nonresonant models.}
\end{table*}

The discrete $P$-wave phase shifts determined for several $\vec{P}, \Lambda$ are listed in
Table \ref{tab:gevp_results} next to the invariant masses. The first uncertainty given is
the statistical uncertainty determined using single-elimination jackknife. The second
uncertainty given is the systematic uncertainty resulting from the choice of $t_{min}$ in
the fits to the GEVP principal correlators; it is computed by repeating the extraction of
$\delta$ with $t_{min}+a$, and then applying Eq.~(\ref{eq:sigmasys}) to the two phase
shift results.

We then fit the models described in Sec.~\ref{sec_about_rho} to the phase shift points.

To correctly estimate the uncertainties of the model parameters, we include the
uncertainties in both $\sqrt{s}$ and $\delta_1$ in the construction of the $\chi^2$
function. To this end, we define
\begin{widetext}
\begin{equation}
 \chi^2 = \sum_{\vec{P},\Lambda,n}\:\: \sum_{\vec{P}^\prime,\Lambda^\prime,n^\prime}\:\: \sum_{i\in \left\{ \sqrt{s_n^{\Lambda, \vec{P}}},\:\: \delta_1(s_n^{\Lambda, \vec{P}}) \right\}}\:\: \sum_{j\in \left\{ \sqrt{s_{n^\prime}^{\Lambda^\prime, \vec{P}^\prime}},\:\: \delta_1(s_{n^\prime}^{\Lambda^\prime, \vec{P}^\prime}) \right\}} (y_i^{avg} - f_i ) [C^{-1}]_{ij}  (y_j^{avg} - f_j ),
\end{equation}
\end{widetext}
where $i$ and $j$ are generalized indices labeling both the data points for $\sqrt{s}$ and
$\delta_1$. The covariance matrix $C$ is therefore a $2N\times 2N$ matrix, where $N=15$ is
the total number of energy levels included in the fit (see the last column of Table
\ref{tab:gevp_results}). For $i$ corresponding to a $\sqrt{s}$ data point, the function
$f_{i}$ is equal to a nuisance parameter $ \sqrt{s_n^{\Lambda, \vec{P}}}$; for $i$
corresponding to a $\delta_1$ data point, the function $f_{i}$ is equal to the phase shift
model evaluated at the corresponding $ \sqrt{s_n^{\Lambda, \vec{P}}}$. The total number of
parameters in the fit is thus equal to $N$ plus the number of parameters in the phase
shift model.

When constructing the covariance matrix, we included the correlations between all
invariant-mass values and the correlations between all phase-shift values. We found that
the covariance matrix becomes ill-conditioned when including also the cross-correlations
between $\sqrt{s}$ and $\delta_1$ as expected when dealing with fully correlated data. We
therefore neglect these contributions in the evaluation of $\chi^2$. The cross-correlations
are nevertheless accounted for in our estimates of the parameter uncertainties, which are
obtained by jackknife resampling.

The fit of the simplest possible model, {\bf BW I}, is shown as the blue curve in
Fig.~\ref{fig:MIvMII} and the resulting parameters $m_\rho$ and $g_{\rho\pi\pi}$ are given
in the first row of Table \ref{tab:d_mIvsmII}. As before, the first uncertainty given is
statistical, and the second uncertainty is the systematic uncertainty arising from the
choice of $t_{min}$. To obtain the latter, we repeated the Breit-Wigner fit for the phase
shifts extracted with $t_{min}+a$ for all energy levels, and then applied
Eq.~(\ref{eq:sigmasys}) to $m_\rho$ and $g_{\rho\pi\pi}$. We follow the same procedure for
all other models.

We then investigate the effect of adding the Blatt-Weisskopf barrier factors
\cite{VonHippel:1972fg} to the decay width appearing in the Breit-Wigner parametrization of $\delta_1(s)$,
which leads to model {\bf BW II}. The resulting fit is shown as the red curve in
Fig.~\ref{fig:MIvMII} (alongside the blue {\bf BW I} curve) and the resulting parameters
are given in the second row of Table \ref{tab:d_mIvsmII}. The {\bf BW II} model appears to
give a slightly better description of the data at high invariant mass, but the paramaters
$m_\rho$ and $g_{\rho\pi\pi}$ are essentially unchanged. Furthermore, the centrifugal
barrier radius $r_0$ is consistent with zero at the $1.1 \sigma$ level, indicating that it
is not a very significant degree of freedom. We note that this could be related to the
high pion mass used in our calculation, which limits the phase space available for the
decay and suppresses the centrifugal barrier effect.

We continue by investigating whether there is a nonresonant contribution to the scattering
phase shift. We first add a nonresonant contribution to the resonant model {\bf BW I}. In
Fig.~\ref{fig:delta_MI} we compare the resonant-only fit (blue curve) with the full fits
for three different forms of the nonresonant contributions (red curves). For clarity we
also show the nonresonant contributions obtained from the full fits separately (orange
curves). The fit results are given in Table \ref{tab:mI_bg}.
We find that the parameters of each of the three parametrizations {\bf NR I} (constant phase),
{\bf NR II} (a nonresonant phase depending linearly on $s$), and {\bf NR III} (zeroth-order
ERE) are consistent with zero, and the results for $m_\rho$ and $g_{\rho\pi\pi}$ also do not
change significantly.

Performing the analoguous analysis for the resonant model {\bf BW II} gives the
phase shift curves shown in Fig.~\ref{fig:delta_MII} and fit parameters in Table \ref{tab:mII_bg}.
Again, the parameters of the nonresonant contribution are consistent with zero, and
$m_\rho$ and $g_{\rho\pi\pi}$ do not change significantly. When adding the nonresonant
contributions to the {\bf BW II} model, the uncertainty of the centrifugal barrier parameter
$r_0$ increases substantially.

Overall, we find that the minimal resonant model {\bf BW I} is sufficient for a good description
of our results for the elastic $I=1$ $\pi\pi$ $P$-wave scattering.

\null

\subsection{\texorpdfstring{Fitting a $t$-matrix to the spectrum}{}}\label{subsec_tfit}

\begin{figure*}
  \includegraphics[width=0.9\textwidth]{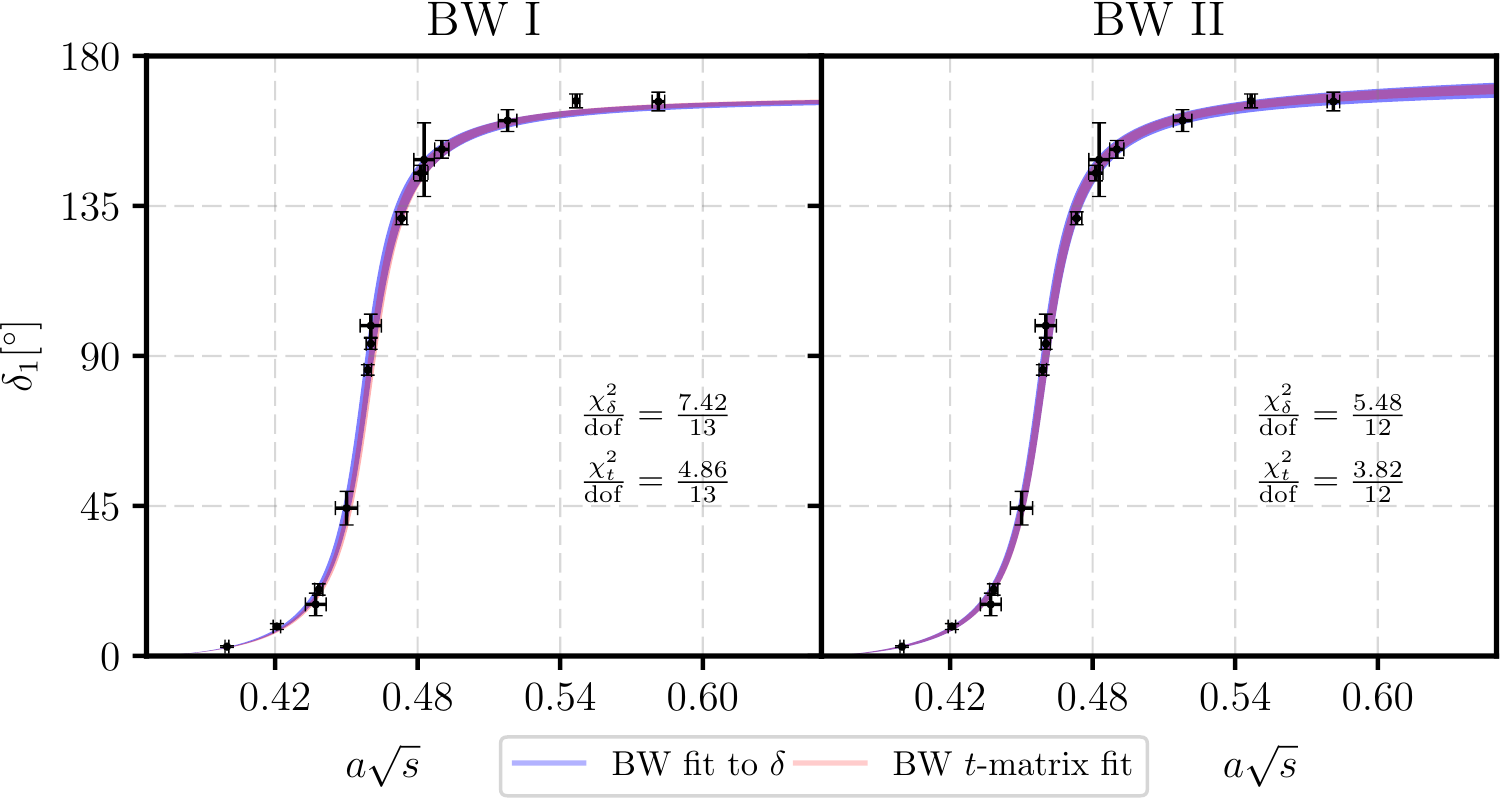}
  \caption{\label{fig:dvT}Comparison of $t$-matrix fit and fit to the phase shifts for Breit-Wigner
   models I and II.}
\end{figure*}

  \begin{table*}
    \begin{tabular}{|l | l | l | l | l|}
      \hline
      Fit type & $\frac{\chi^2}{{\rm dof}}$ & $am_\rho$ & $g_{\rho\pi\pi}$ & $(ar_0)^2$ \cr
      \hline
      {\bf BW I } Fit to $\delta_1$ & $0.571$ & $0.4599(19)(13) $  &  $ 5.76(16)(12) $ & \cr
      {\bf BW I } $t$-matrix fit & $0.374$ &  $0.4609(16)(14) $  &  $5.69(13)(16) $ & \cr
      {\bf BW II } Fit to $\delta_1$ & $0.457$ & $0.4600(18)(13) $  &  $ 5.79(16)(12) $  &  $8.6(8.0)(1.2) $\cr
      {\bf BW II } $t$-matrix fit    & $0.318$ & $0.4603(16)(14)$   & $5.77(13)(13)$     &  $9.6(5.9)(3.7)$\cr
      \hline
    \end{tabular}
    \caption{\label{tab:dvT}Comparison of $t$-matrix fits with direct fits to the phase shifts.}
  \end{table*}

For the $t$-matrix fit to the spectrum, we define the $\chi^2$ function as
\begin{widetext}
\begin{align}
\label{eq:tmatchi2}
\chi^2 = \sum_{\vec{P},\Lambda,n}\:\: \sum_{\vec{P}^\prime,\Lambda^\prime,n^\prime}\:\: \bigg( \sqrt{s_n^{\Lambda, \vec{P}}}^{[avg]} - \sqrt{s_n^{\Lambda, \vec{P}}}^{[model]} \bigg) [C^{-1}]_{\vec{P},\Lambda,n;\vec{P}',\Lambda',n'}\bigg( \sqrt{s_{n'}^{\Lambda', \vec{P}'}}^{[avg]} - \sqrt{s_{n'}^{\Lambda', \vec{P}'}}^{[model]}  \bigg),
\end{align}
\end{widetext}
where the invariant-mass values $\sqrt{s_{n'}^{\Lambda', \vec{P}'}}^{[model]}$ are
obtained by solving the inverse L\"uscher problem, i.e. determining the finite-volume
spectrum from a given $t$-matrix model \cite{Guo:2012hv,Dudek:2012xn}. Above, $C$ is the
matrix of covariances between all invariant-mass values labeled by $\vec{P},\Lambda,n$ (in
our case, this is a $15\times15$ matrix). The only fit parameters in this approach are the
parameters of the $t$ matrix (for example, $am_\rho$ and $g_{\rho\pi\pi}$ for the {\bf BW
I} model).

When fitting the $t$-matrix directly to the spectrum we consider only the two resonant
models, as results from Sec.~\ref{subsec_directfit} show no indication of a nonresonant
phase contribution. The parameters obtained from the $t$-matrix fits are compared to the
parameters of the direct fits to the phase shifts in Table \ref{tab:dvT}. The plots of the
models with parameters from the two different fit approaches are compared in
Fig.~\ref{fig:dvT}. The central values and uncertainties obtained with the two methods are
consistent, which confirms previous findings \cite{Guo:2012hv,Dudek:2012xn} that the two
approaches are equivalent not only theoretically but also in practice. We note that the
values of $\chi^2/{\rm dof}$ are generally quite small. We have tested for the presence of
autocorrelations in the data using binning, but found no significant effect.

\subsection{\texorpdfstring{Final result for the $\rho$ resonance parameters}{}}\label{subsec_result}

\begin{figure}
  \includegraphics[width=\columnwidth]{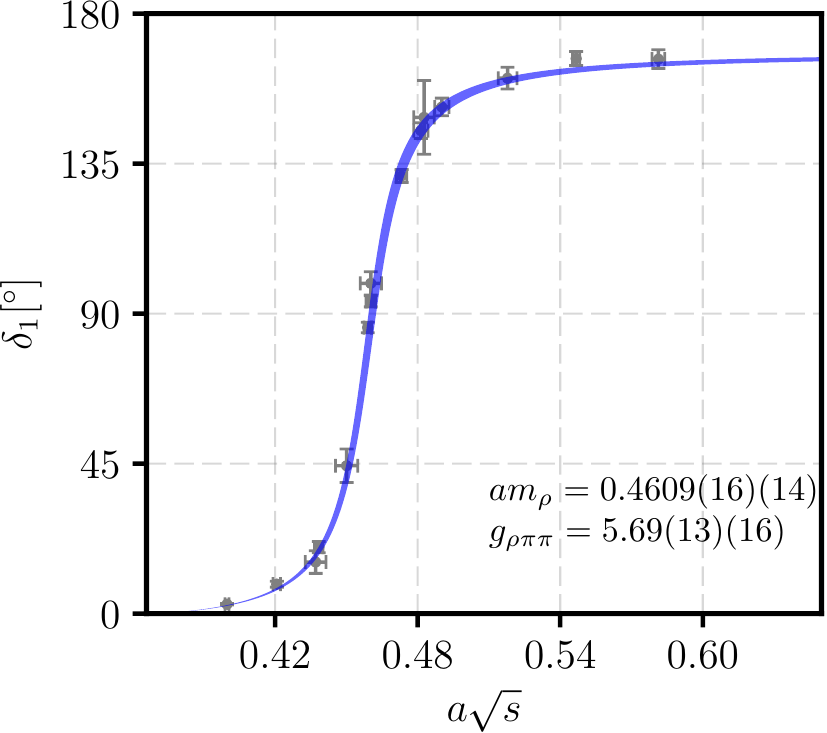}
  \caption{\label{fig:t_pipi_MI} Final result of fitting the resonant model {\bf BW I} to
   the spectrum via the $t$-matrix fit. The gray data points are the results of the
   individual phase shift extractions for each energy level, and are not used in the
   $t$-matrix fit.}
\end{figure}

Given the discussion in the previous sections, we choose to quote the results of the
$t$-matrix fit with the resonant Breit-Wigner model ${\bf BW I}$ as our final values of
$am_{\rho}$ and $g_{\rho\pi\pi}$ for the ensemble of gauge configurations used here [with
$am_{\pi}=0.18295(36)$ and $am_{N}=0.6165(23)$]:
\begin{align}
\label{eq:fin_res}
\bordermatrix{ & & & \cr
  am_{\rho}  = 0.4609(16)(14)       &    1.0    &   0.326  \cr
  g_{\rho\pi\pi}  = 5.69(13)(16)   &           &   1.0    \cr}\ .
\end{align}
The phase shift curve of our chosen fit is shown in Fig.~\ref{fig:t_pipi_MI}.
Above, the first uncertainties given are statistical, and the second uncertainties are the
systematic uncertainties related to the choice of $t_{min}$ in the spectrum analysis. Also
given in Eq.~(\ref{eq:fin_res}) is the statistical correlation matrix for $am_{\rho}$ and
$g_{\rho\pi\pi}$. The exponentially suppressed finite-volume errors in $m_{\rho}$ and
$g_{\rho\pi\pi}$ are expected to be of order ${\cal O}(\E^{-m_{\pi}L})\approx0.3\%$. Given
that we have only one lattice spacing, we are unable to quantify discretization errors
(except in the pion dispersion relation, Sec.~\ref{sec_pion}, where we find $c^2$
to be consistent with $1$ within $2\%$). Using the lattice spacing determined from the
$\Upsilon(2S)-\Upsilon(1S)$ splitting (see Table \ref{tab:lattice}), we obtain
\begin{align}
 m_\pi &= 316.6(0.6)_{stat}(2.1)_{a}\:\:{\rm MeV}, \nonumber \\
 m_\rho &= 797.6(2.8)_{stat}(2.4)_{sys}(5.4)_{a}\:\:{\rm MeV},  \nonumber \\
 g_{\rho\pi\pi}  &= 5.69(13)_{stat}(16)_{sys}.
\end{align}

It is important to note
that the lattice spacing uncertainty given here is statistical only. As a consequence of the heavier-than-physical pion mass
and lattice artefacts, different quantities used to set the scale of an individual ensemble
yield different results for $a$ and hence for $m_\pi$ and $m_\rho$ in units of MeV.
We therefore prefer to report the dimensionless ratios
\begin{align}
 \frac{a m_\pi}{a m_N} &= 0.2968(13)_{stat}, \nonumber \\
 \frac{a m_\rho}{a m_N} &= 0.7476(38)_{stat}(23)_{sys},
\end{align}
in which the lattice scale cancels.

\begin{figure*}
  \includegraphics[width=2.05\columnwidth]{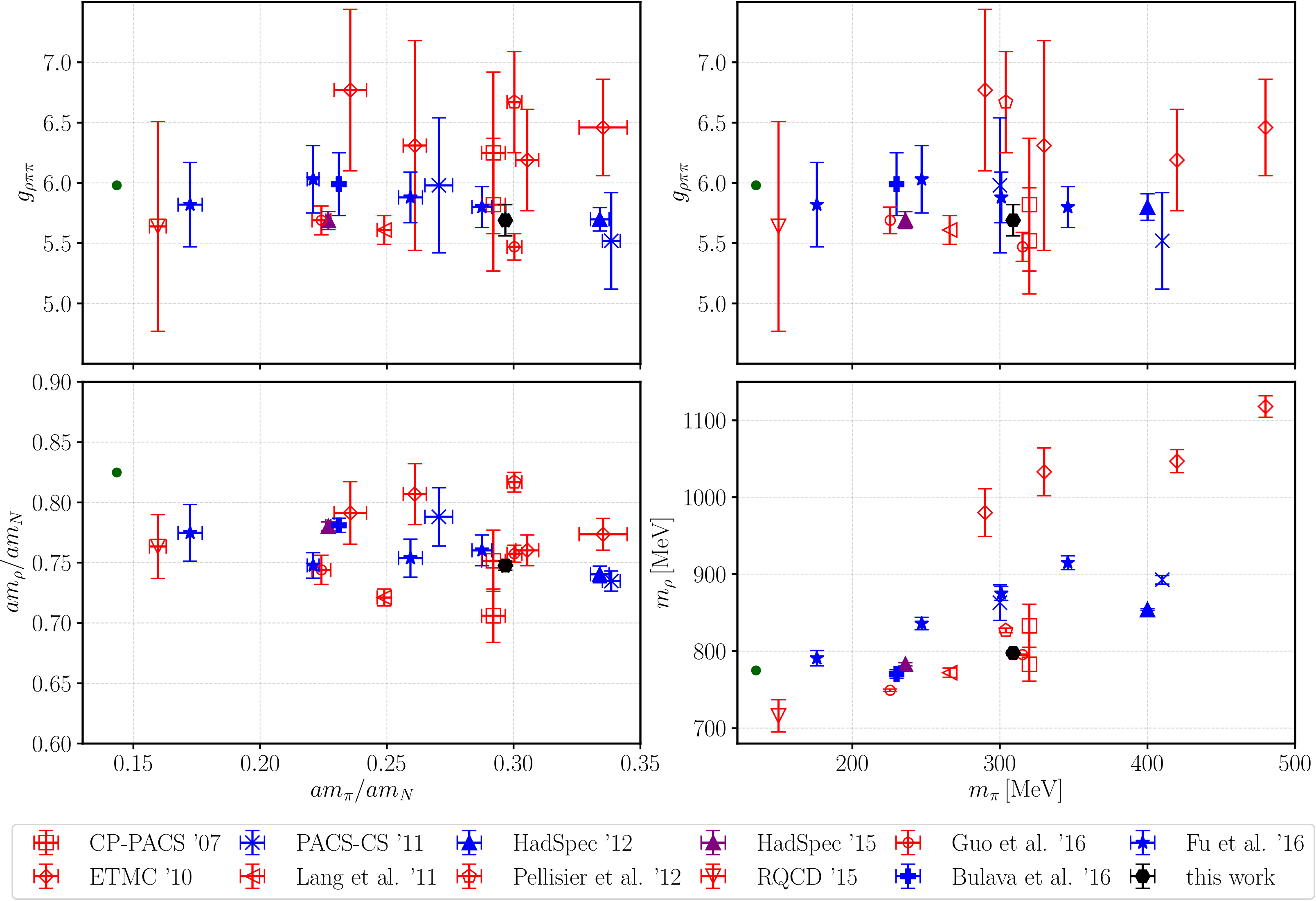}
  \caption{\label{fig:prev_comparison} Comparison of our results for the $\rho$ mass and coupling with previous lattice QCD calculations.
  In the two left panels, we use the dimensionless ratios $am_\rho/am_N$ and $am_{\pi}/am_N$, while in the two right panels
  we use $m_\pi$ and $m_\rho$ in MeV as reported by each collaboration (with different scale setting methods; the error bars do not include the
  scale-setting ambiguities). The open red symbols mark calculations with $N_f=2$ gauge ensembles, while the filled blue symbols denote
  calculations with $N_f=2+1$ sea quarks; the only study so far that explicitly included the $K\bar{K}$ channel,
  HadSpec '15, is presented as a purple upward facing triangle. The results of our present work are shown
  with filled black hexagons. In the left-hand plots, the HadSpec '15 results are offset horizontally by $-1.8\%$ so that they do not
  overlap with the result of Bulava et al.~'16. In the right-hand plots, we offset our results by $-8\;{\rm MeV}$ to avoid overlap with Guo et al.~'16.
  The experimental values \cite{Olive:2016xmw}, where $g_{\rho\pi\pi}$ was calculated from
  $\Gamma$ using Eq.~(\protect\ref{eq:Gamma_Pwave}), are shown with filled green circles.}
\end{figure*}

In Fig.~\ref{fig:prev_comparison} we compare our results for the $\rho$ coupling and mass with the results of previous studies
performed by the CP-PACS collaboration (CP-PACS '07) \cite{Aoki:2007rd}, the ETMC collaboration (ETMC '10) \cite{Feng:2010es},
the PACS-CS collaboration (PACS-CS '11) \cite{Aoki:2011yj}, Lang et al. (Lang et al. '11) \cite{Lang:2011mn},
the Hadron Spectrum collaboration (HadSpec '12 and HadSpec '15) \cite{Dudek:2012xn,Wilson:2015dqa},
Pellisier et al. (Pellisier et al. '12) \cite{Pelissier:2012pi},
the RQCD collaboration (RQCD '15) \cite{Bali:2015gji}, Guo et al. (Guo et al. '16) \cite{Guo:2016zos},
Bulava et al. (Bulava et al. '16) \cite{Bulava:2016mks}, and Fu et al. (Fu et al. '16) \cite{Fu:2016itp}.
In the right half of the figure, we use the values of $m_\pi$ and $m_\rho$ in MeV as reported in each reference.
In the left half of the figure, we instead use the dimensionless ratios $a m_\pi/a m_N$ and $a m_\rho/a m_N$,
where $a m_\pi$ and $a m_N$ are the pion and nucleon masses in lattice units computed on the same ensemble as $a m_\rho$.
The nucleon masses were obtained from Refs.~\cite{Namekawa:2004bi, Alexandrou:2010hf, Lin:2008pr, Detmold:2015qwf, Lang:2012db,
Aoki:2008sm, Bali:2016lvx, MILCmN}.

We find that our value for the coupling $g_{\rho\pi\pi}$ is in good agreement with previous studies both as a
function of $m_{\pi}$ and $a m_{\pi}/am_N$. Furthermore, it is
consistent with the general finding that $g_{\rho\pi\pi}$  has no discernible pion-mass
dependence in the region between $m_{\pi,phys}$ and approximately  $3\,m_{\pi,phys}$.

Concerning the results for the $\rho$ mass, the left and right panels of Fig.~\ref{fig:prev_comparison} show very different
behavior. This discrepancy arises from the different methods used to set the lattice scale on a single ensemble, which can lead to
misleading conclusions. To avoid the substantial ambiguities associated with the scale setting,
we only consider the dimensionless ratio $am_\rho/am_N$ in the following discussion.

The $N_f=2+1$ results for $am_\rho/am_N$ obtained with Wilson-Clover-based fermion actions all approximately
lie on a straight line leading to the experimental value (shown as the filled green circle in Fig.~\ref{fig:prev_comparison}).
The $N_f=2+1$ data points using staggered fermions (Fu et al. '16) are consistent with that line except for one outlier.

The $N_f=2$ results are dispersed around the $N_f=2+1$ values in both directions. The discrepancies between
the different results could arise from any of several systematic effects, such as excited-state contamination in the determination of the $\pi\pi$
spectrum or the nucleon mass, various potential issues in fitting the data, and discretization errors which manifest
themselves for example in deviations from the relativistic continuum dispersion relation for the single-pion energies.
Additionally, the L\"uscher method only addresses power-law finite volume effects
and does not take into account the exponentially suppressed finite-volume effects which are estimated to
scale asymptotically as $O(e^{-m_{\pi}L})$. Note that for some of the studies, these can be as high as
$O(10\%)$ and it is thus not clear whether the asymptotic regime is reached.
An example for systematics associated with the pion dispersion relation
can be seen in the CP-PACS '07 study, where the two different results for $a m_\rho$ at the same pion mass were
obtained using either the relativistic continuum dispersion relation or a free-boson lattice dispersion relation.
An example of systematic effects that might be associated with the data analysis
can be seen when comparing the Pellisier et al.~'12 results with the
Guo et al.~'16 results at $am_{\pi}/am_{N} \approx 0.3$. Both studies used the same ensemble, but
arrive at significantly different values for the $\rho$ resonance parameters.

Keeping these caveats in mind, it is nevertheless interesting to note that our
$N_f=2+1$ results for both $am_\rho/am_N$ and $g_{\rho\pi\pi}$ agree well with
the recent $N_f=2$ results from Guo et al.~'16 at almost the same pion mass.
This suggests that the effects of the dynamical strange quark are small at
$m_\pi\approx 320$ MeV. The HadSpec `15 study, which explicitly included the $K\bar{K}$ channel
in their valence sector, provides further evidence
that the strange quark does not play a major role in the $\rho$ resonance mass.

\section{Summary and Conclusions}\label{sec_conclusions}

We have presented a $(2+1)$-flavor lattice QCD calculation of $I=1$, $P$ wave $\pi\pi$ scattering
at a pion mass of approximately 320 MeV. The calculation was performed in a large volume of
$(3.6\:{\rm fm})^3\times (10.9\:{\rm fm})$ and utilized all irreps of $LG(\vec{P})$
with total momenta up to $|\vec{P}| \leq \sqrt{3} \frac{2\pi}{L}$. Using a method
based on forward, sequential, and stochastic propagators that scales well with the
volume, we have achieved high statistical precision ($0.35\%$ for $a m_{\rho}$ and $2.3\%$ for $g_{\rho\pi\pi}$).

We compared two different methods to determine the energy spectrum: the generalized
eigenvalue problem (GEVP), and multi-exponential direct matrix fits to the correlation matrices (MFA).
A careful investigation of the dependence on the fit ranges showed that both approaches
are equally powerful and give consistent results.

After determining the elastic scattering phase shifts from the spectrum, we analyzed
several different models for the energy dependence of the $\pi\pi$ scattering amplitude.
We investigated two different Breit-Wigner forms, one with added Blatt-Weisskopf barrier factors,
and found that the addition of this degree of freedom was not necessary to describe our data.
This could be due to the higher-than-physical pion mass used in this work. Additionally,
we examined whether there is a nonresonant contribution to the scattering phase shift,
finding that it is consistent with zero within our statistical uncertainties.

Regarding the technical aspects of the analysis, we also compared two different ways of determining
the scattering parameters: extracting the discrete phase shift points from each individual energy level
(which is only feasible for elastic scattering) versus fitting the parameters of the $t$-matrix directly
to the spectrum (as is also done in multichannel studies). We have demonstrated numerically
that both methods are equivalent.

In summary, we found that the $I=1$, $P$-wave $\pi\pi$ scattering at $m_\pi\approx 320$ MeV
is well described in the elastic energy region by the minimal resonant Breit-Wigner model {\bf BW I} (defined in
Sec.~\ref{sec_about_rho}) with the parameters given in Eq.~(\ref{eq:fin_res}). A comparison
with previous lattice results, shown in Fig.~\ref{fig:prev_comparison}, revealed that (i)
it is important to use dimensionless ratios such as $am_\rho/am_N$ and $am_\pi/am_N$
to avoid scale setting ambiguities, and (ii) there are signs of significant systematic errors
whose origins are difficult to disentangle without additional dedicated calculations.

\acknowledgments

We are grateful to Kostas Orginos for providing the gauge field ensemble, which was generated using
resources provided by XSEDE (supported by National Science Foundation Grant No.~ACI-1053575).
We thank Raul Brice\~no, Sean Fleming, Doug Toussaint, and Bira Van Kolck for valueable
discussions. SM and GR are supported by National Science Foundation Grant No.~PHY-1520996;
SM and SS also acknowledge support by the RHIC Physics Fellow Program of the RIKEN BNL Research Center.
JN and AP were supported in part by the U.S. Department of Energy Office of Nuclear Physics
under Grant Nos.~DE{}-SC-0{}011090 and DE-{}FC02-06ER41444.
We acknowledge funding from the European Union's Horizon 2020 research and innovation programme
under the Marie Sklodowska-Curie grant agreement No 642069. S.~P.~is a Marie Sklodowska-Curie fellow supported by the
the HPC-LEAP joint doctorate program.
This research used resources of the National Energy Research Scientific Computing Center,
a DOE Office of Science User Facility supported by the Office of Science of the U.S.
Department of Energy under Contract No.~DE-AC02-05CH11231. The computations were performed
using the Qlua software suite \cite{QLUA}.





\end{document}